\documentclass[aps,superscriptaddress,twocolumn]{revtex4}
\usepackage{amsmath}
\usepackage{graphicx}
\usepackage{dcolumn}
\usepackage{verbatim}
\usepackage{times}
\usepackage{subfigure}
\usepackage{bm}
\usepackage{color}
\usepackage[colorlinks,citecolor=blue,linkcolor=blue,dvipdfm]{hyperref}
\usepackage{subfigure}

\usepackage{booktabs}
\usepackage{appendix}

\begin{document}

\title{Phase evolution of Ce-based heavy-fermion superconductors under compression: a combined first-principle and effective-model study}

\author{Hao-Tian Ma\footnote{These authors contribute equally to this work.}}
\affiliation{College of Physics and Electronic Information Engineering, Guilin University of Technology, Guilin 541004, China}
\author{Peng-Fei Tian\footnote{These authors contribute equally to this work.}}
\affiliation{College of Physics and Electronic Information Engineering, Guilin University of Technology, Guilin 541004, China}
\author{Da-Liang Guo}
\affiliation{College of Physics and Electronic Information Engineering, Guilin University of Technology, Guilin 541004, China}
\author{Xing Ming}
\affiliation{College of Physics and Electronic Information Engineering, Guilin University of Technology, Guilin 541004, China}
\author{Xiao-Jun Zheng}
\affiliation{College of Physics and Electronic Information Engineering, Guilin University of Technology, Guilin 541004, China}
\author{Yu Liu}
\affiliation{Laboratory of Computational Physics, Institute of Applied Physics and Computational Mathematics, Beijing 100088, China}
\author{Huan Li}
\email{lihuan@glut.edu.cn}
\affiliation{College of Physics and Electronic Information Engineering, Guilin University of Technology, Guilin 541004, China}

\date{\today}

\begin{abstract}

In many Ce-based superconductors, superconducting (SC) phases emerge or can be tuned in proximity to the antiferromagnetic (AF) quantum critical point (QCP), but so far the explicit phase evolution near the QCP lack theoretical understanding. Here, by combing the density functional theory plus dynamical mean-field theory (DFT+DMFT) with effective-model calculations, we provide a theoretical description for Ce-based superconductors under compression. DFT+DMFT calculations for the normal states reveal that the Kondo hybridizations are significantly enhanced under compression, while the initially localized $f$ electrons become fully itinerant via localized-itinerant crossover. We then construct an effective model and show that with the extracted Kondo coupling and RKKY exchange strengths from first-principle calculations, the ground-state phases of these materials can be properly predicted.
We also show that the coexistence of magnetic correlation and Kondo hybridization can drive AF+SC coexisting state in narrow compression region.
Under compression, competition between Kondo and RKKY interactions can drive successive transitions, from AF phase to AF+SC coexisting phase, then to paramagnetic SC phase via an AF transition which generates the QCP, and finally to normal Kondo paramagnetic (KP) phase through an SC-KP transition induced by the localized-itinerant crossover. Our study gives proper explanation to the pressure-induced QCP and SC-KP transition, and to the phase evolution in pressured Ce-based superconductors, and can help to understand the SC states around the ferromagnetic quantum transition points in uranium-based superconductors.

\end{abstract}

\maketitle

\section{Introduction}

Heavy-fermion (HF) materials are characterized by Kondo hybridization between conduction and $f$ electrons, in which the $f$ electrons are correlated via Coulomb repulsion and tend to form local moments. At low temperature, the Kondo screening of $f$ electrons by conducting electrons induces many-body Kondo singlet state, resulting in strong enhancement of quasi-particle mass and intense Kondo resonance peak in the density of states (DOS) near the Fermi level~\cite{Andres75}. In some HF materials such as CeCu$_2$Si$_2$, CeMIn$_5$ (M=Rh, Co, Ir) and recently discovered CeRh$_2$As$_2$ and CeSb$_2$, superconducting (SC) phases have been observed with large coefficient of specific heat, illustrating the emergence of heavy-fermion superconductivity~\cite{White15,Pfleiderer09,Steglich79,Yuan03,Sidorov02,Ida08,Shang14,Khim21,Squire22}.
In these heavy-fermion SC compounds, rich phases and phase transitions are observed under pressure. At ambient pressure, CeCu$_2$Si$_2$, CeRhIn$_5$ and CeSb$_2$ are antiferromagnetically ordered, then they enter into
SC phases when AF orders are gradually suppressed at higher pressure, showing up an arc-shaped SC critical temperature $T_c$ surrounding the AF quantum critical point (QCP) in their
pressure-temperature phase diagrams~\cite{Yuan03,Ida08,Squire22}. In intermediate pressure region in CeCu$_2$Si$_2$, CeRhIn$_5$ and CeSb$_2$, etc, AF orders can coexist with heavy-fermion superconductivity, evidencing the emergence of AF+SC coexisting phase~\cite{Kawasaki08,Pagliuso01,Nair10,Park06,Sidorov13,Movshovich96}. By contrast, for CeRh$_2$As$_2$, the ground state is already AF+SC phase at ambient pressure, and its $T_c$ gradually decreases with enhanced pressure~\cite{Siddiquee22,Kibune22}. While for CeCoIn$_5$ and CeIrIn$_5$, at ambient pressure, the SC phases emerge without AF long-range order~\cite{Sidorov02,Shang14}.

The vicinity of many typical Ce-based heavy-fermion SC compounds to AF orders establishes that magnetic correlations between $f$ electrons play an important role in the development of superconductivity, in particular, the occurrence of superconductivity near the pressure-induced magnetic QCP in CePd$_2$Si$_2$ and CeCu$_2$Si$_2$ supports a SC pairing mediated by AF correlations~\cite{Andrei89,Millis93,Mathur98,Arndt11,Park06,Scalapino12,Dyke14}. In addition, the majority of heavy-fermion SC compounds preserve space-inversion symmetry, ensuring that the Cooper pairs arise in spin-singlet even-parity channel~\cite{Akbari12,An10}. In spite of these theoretical investigations, the microscopic mechanism governing the coexistence of heavy-fermion SC with AF order at intermediate pressure region remains lack of explicit theoretical explanation so far. In particular, the successive phase evolution with applied pressure in heavy-fermion SC materials, and why they exhibit distinct ground-state phases at ambient pressure are still lack of deep understanding beyond the phenomenological level~\cite{Yang15,Yang152}.

In this article, we systematically explore the effect of pressure on typical Ce-based heavy-fermion SC materials through density functional theory combing with dynamical mean-field theory (DFT+DMFT), in company with effective-model description.
Firstly, by DFT+DMFT calculations for the normal states of these materials at low temperature, we show that volume compression can significantly enhance the Kondo hybridization between conduction and $f$ electrons, meanwhile weaken the local-moment degree of $f$ electrons, then eventually drive a localized-to-itinerant crossover of the Ce-4$f$ states at rather high volume compression ratio. Based on these DFT+DMFT results, we construct an effective Kondo-Heisenberg lattice model and successfully derive an AF+SC coexisting phase in which the SC pairing is mediated by short-range singlet pairing between $f$ electrons in the context of $c$-$f$ hybridization and AF long-range order. By including the pressure variation of Kondo coupling $J_K$ and Ruderman-Kittel-Kasuya-Yosida (RKKY) superexchange strength $J_H$ extracted via first-principle calculations, we find that in the context of competition between $J_K$ and $J_H$, the increasing pressure can drive successive phase transitions, from AF ordered phase to AF+SC coexisting phase, then to paramagnetic heavy-fermion SC phase after an AF transition, and finally to Kondo paramagnetic (KP) phase through an SC transition, in which the AF transition can be related to the QCP in heavy-fermion SC materials, while the SC transition is induced by localized-to-itinerant crossover. Such phase-evolution process gives a qualitatively explanation to the experimental phase diagrams of heavy-fermion SC compounds, such as CeSb$_2$, CeRhIn$_5$ and CeCu$_2$Si$_2$ under pressure~\cite{Yuan03,Ida08,Squire22}. In addition, we show that the ground-state phases of a variety of Ce-based SC compounds at ambient pressure can be predicted properly according to their degrees of RKKY and Kondo coupling strengths. Furthermore, the localized-to-itinerant crossover in CeRh$_2$As$_2$ under compression provides a theoretical predication for SC transition at higher pressure which may be verified by future experiments.
Our effective-model studies plus DFT+DMFT calculations eventually give an appropriate description for typical Ce-based heavy-fermion SC compounds, to their phases and phase evolutions under ambient and higher pressures.

The rest of this paper is arranged as following. In Sec. II, we will perform DFT+DMFT simulations of typical Ce-based superconductors under compression, through synthetical analyses of the self-energy, impurity hybridization function, DOS, momentum-resolved spectral function and spin susceptibility, we will estimate the strength of Kondo hybridization and degree of $f$ localization at ambient pressure, examine their variation tendency with pressure, through which we identify the localized-to-itinerant crossover at certain compression percentage. In Sec. III, we will construct a minimal effective model, and perform mean-field calculation to derive the AF+SC coexisting phase, then by combining the DFT+DMFT and DFT+U results of Kondo coupling and RKKY strengths with effective model, we give predictions of Ce-based superconductors for their ground-state phases and phase evolutions from ambient pressure.
Eventually, our results by combing first-principle simulations with model calculations successfully depict a qualitative phase diagram of heavy-fermion SC compounds under ambient and increasing pressure. Sec. IV will give a brief conclusion and discussion.

\begin{figure}[tbp]
\hspace{0cm} \includegraphics[totalheight=4.2in]{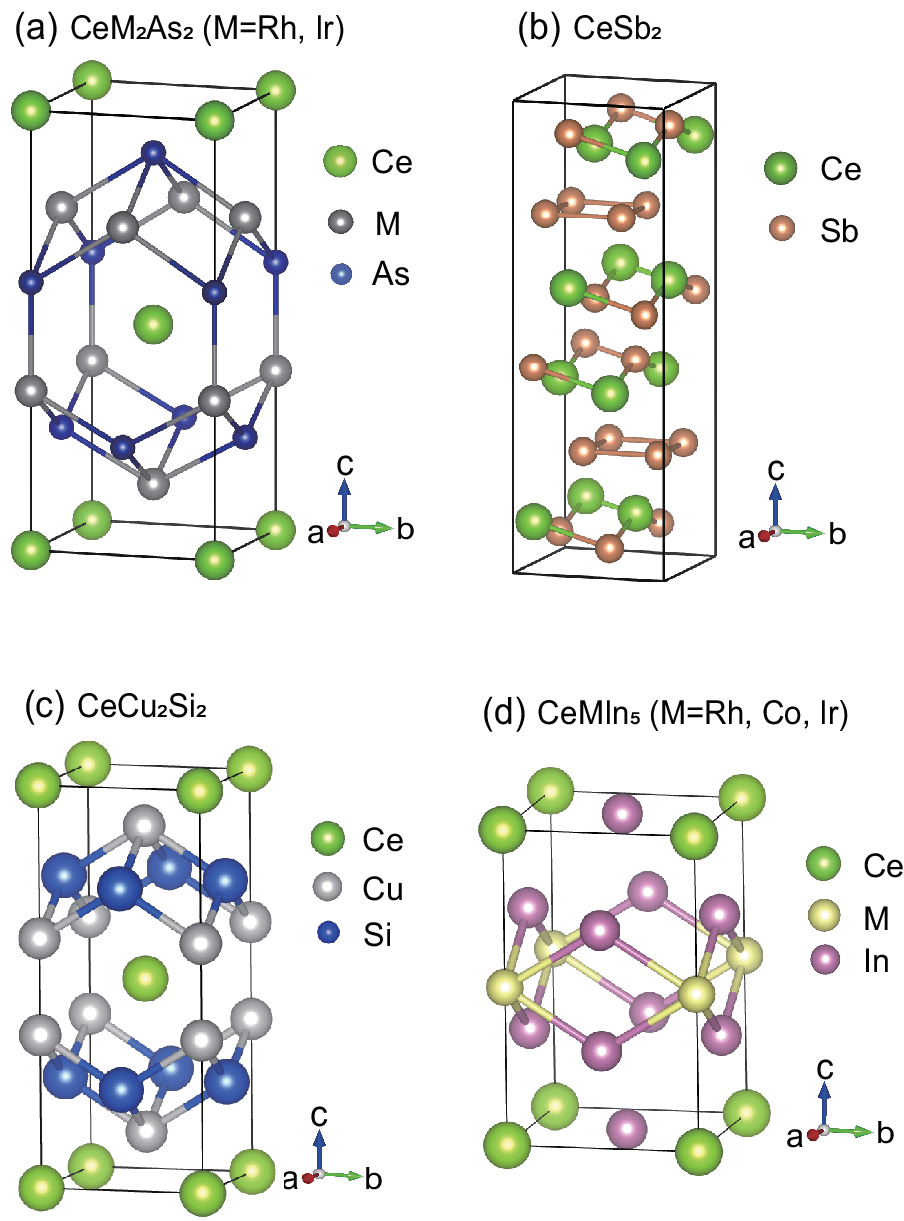}
\centering
\caption{Lattice structures of Ce-based superconductors. These compounds are all crystalized with global inversion symmetry. In (a), the local centrosymmetries of Ce sites in CeRh$_2$As$_2$ and CeRh$_2$Ir$_2$ with tetragonal CaBa$_2$Ge$_2$ type structure are broken~\cite{Ma23}. On $\mathbf{a}$$\mathbf{b}$ plane, the Ce atoms in (a), (c) and (d) form a square-lattice structure.
}
\label{lattice}
\end{figure}

\section{DFT+DMFT simulations and Localized-itinerant crossover}

We adopt the DFT+DMFT method built up in EDMFT code package~\cite{Haule10} to explore the pressure effect in recently discovery heavy-fermion SC compounds CeRh$_2$As$_2$ and CeSb$_2$~\cite{Khim21,Squire22}, of which the lattice structures are shown in Fig.\ref{lattice}. In DFT+DMFT simulation, the DFT part is implemented by full-potential linear augmented plane-wave method embodied in WIEN2k package, and the generated single-particle Kohn-Sham Hamiltonian $\hat{H}_{\mathrm{KS}}$ is combined with an interacting term $\hat{H}_{\mathrm{int}}$ which includes on-site Coulomb repulsion $U$ and Hund's coupling $J$ on Ce-4$f$ electrons, together with a double-counting term $\Sigma_{\mathrm{dc}}$ for self-energy, then the constructed lattice model $\hat{H}_{\mathrm{DFT+DMFT}}=\hat{H}_{\mathrm{KS}}+\hat{H}_{\mathrm{int}}-\Sigma_{\mathrm{dc}}
$ is solved within single-site DMFT algorithm, in which the states within [-10, 10] eV from Fermi level are projected into the Anderson impurity problems. We use nominal double-counting $\Sigma_{\mathrm{dc}}=U(n_f-1/2)-J/2(n_f-1)$, and continuous-time quantum Monte Carlo method (CTQMC) as impurity solver. In order to obtain real-frequency self-energy, we use maximum-entropy method to perform analytical continuation of the output imaginary-frequency self-energy.

\begin{figure*}[tbp]
\hspace{-0cm} \includegraphics[totalheight=2.3in]{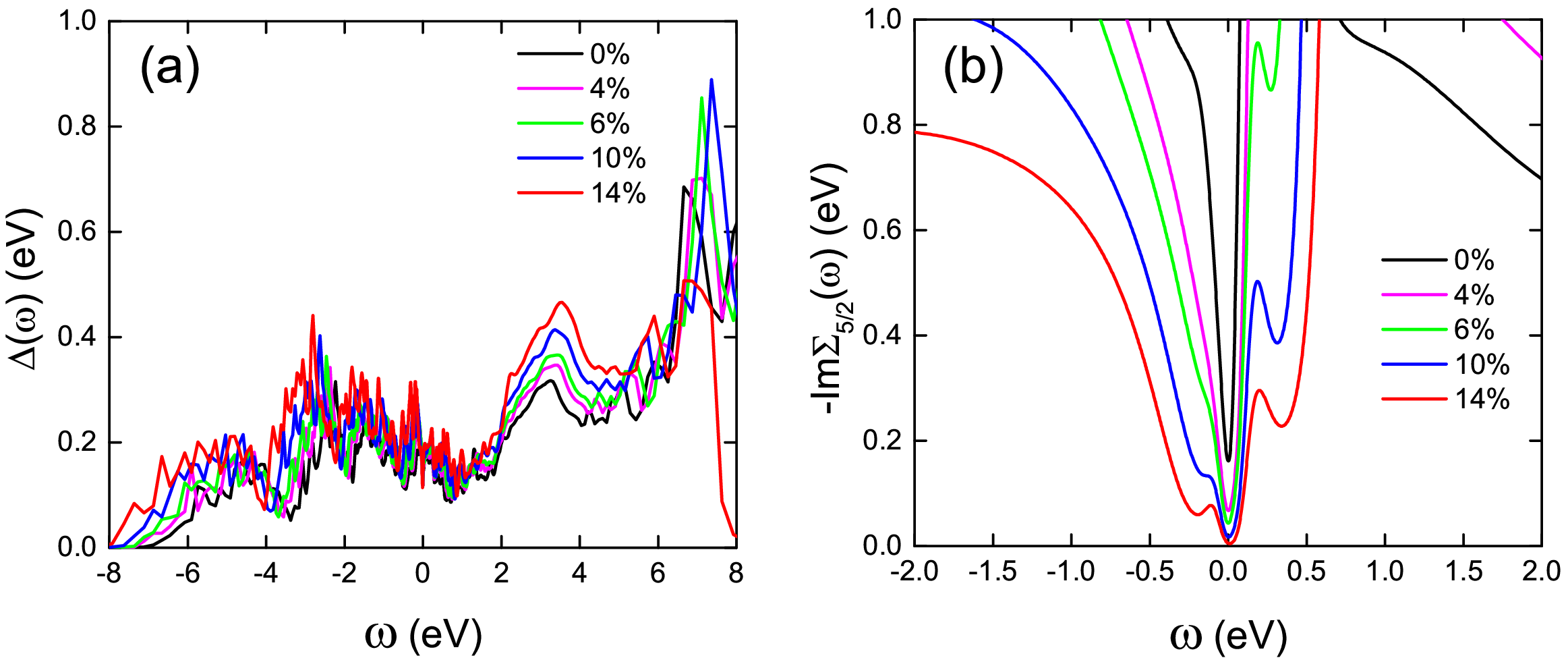}
\caption{
(a) Hybridization function $\Delta(\omega)$ and (b) imaginary part of real-frequency self-energy $-\mathrm{Im}\Sigma_{5/2}(\omega)$ of CeRh$_2$As$_2$ at various compression ratios, all at temperature $T=20$ K. Upon increase of compression ratio, the zero-frequency self-energy $-\mathrm{Im}\Sigma_{5/2}(0)$ is greatly reduced, while the hybridization function is enhanced in wide energy range.
}
\label{SigmaDelta}
\end{figure*}

\begin{figure*}[tbp]
\hspace{-0cm} \includegraphics[totalheight=2.4in]{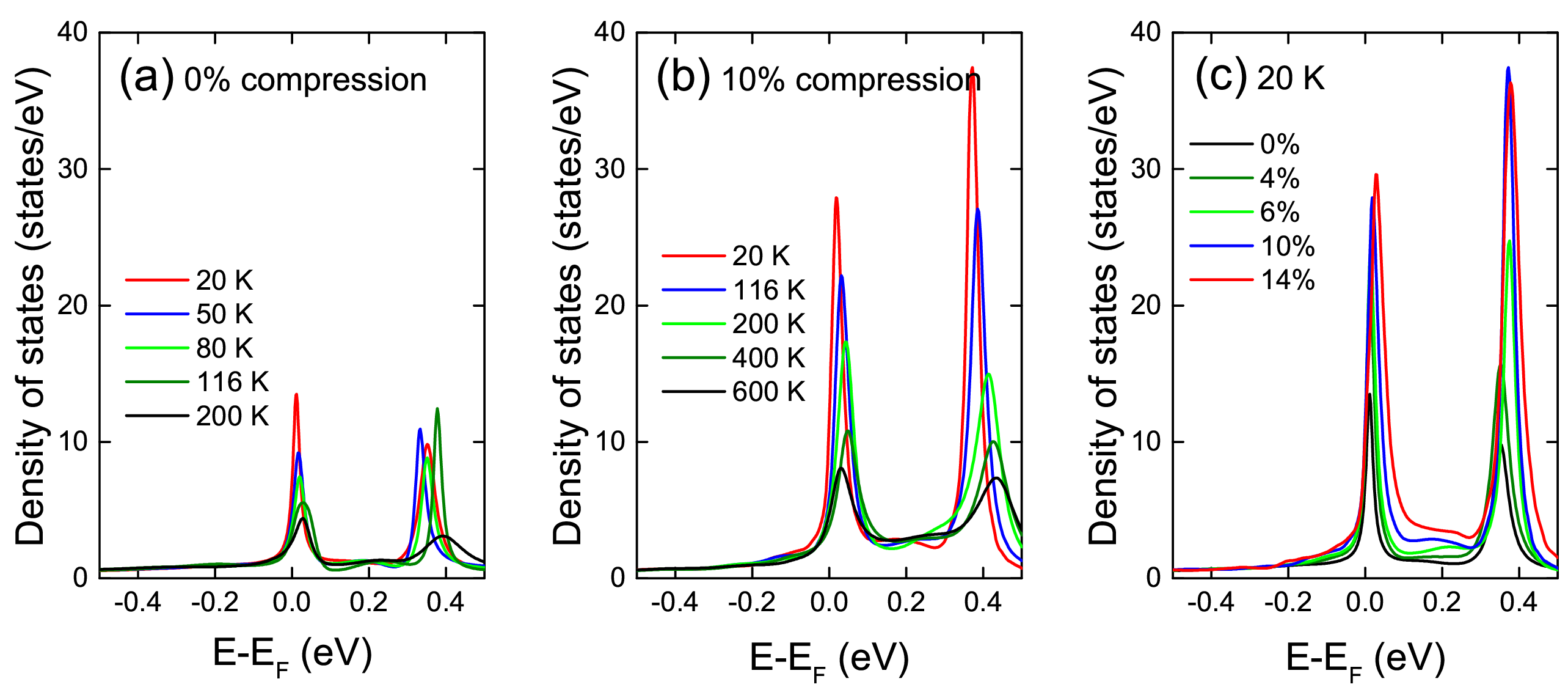}
\caption{
Ce-4$f$ density of states of CeRh$_2$As$_2$ at various temperatures and compression ratios. (a) and (b) show the DOS at 0$\%$ and 10$\%$ volume compression, respectively, indicating the gradual formation of Kondo resonance peak with decreasing temperature. (c) illustrates the enhancement of Kondo resonance peak with increasing compression ratio at 20 K.
}
\label{DOS}
\end{figure*}

In DFT part, we use around 2000 $\mathbf{k}$-points in the Brillouin zone integration (16$\times$16$\times$7 and 20$\times$20$\times$4 $\mathbf{k}$-mesh for CeRh$_2$As$_2$ and CeSb$_2$, respectively), and spin-orbital coupling (SOC) is included throughout the calculations. For CeRh$_2$As$_2$ and CeSb$_2$, typical value of $U$=5.5 eV and $J$=0.7 eV are used for Ce-4$f$ orbits~\cite{Zhu20,Nam19}. In each CTQMC calculation, we use 128 CPU cores to run (3$\sim$10)$\times 10^8$ QMC steps, from temperature 800 K to 20 K. The DFT+DMFT simulations are performed iteratively to reach full-charge self consistence within 50 iterations, then additional five iterations are executed to average the outputs in order to reduce numerical noise.

\begin{figure*}[tbp]
\hspace{-0cm} \includegraphics[totalheight=4.9in]{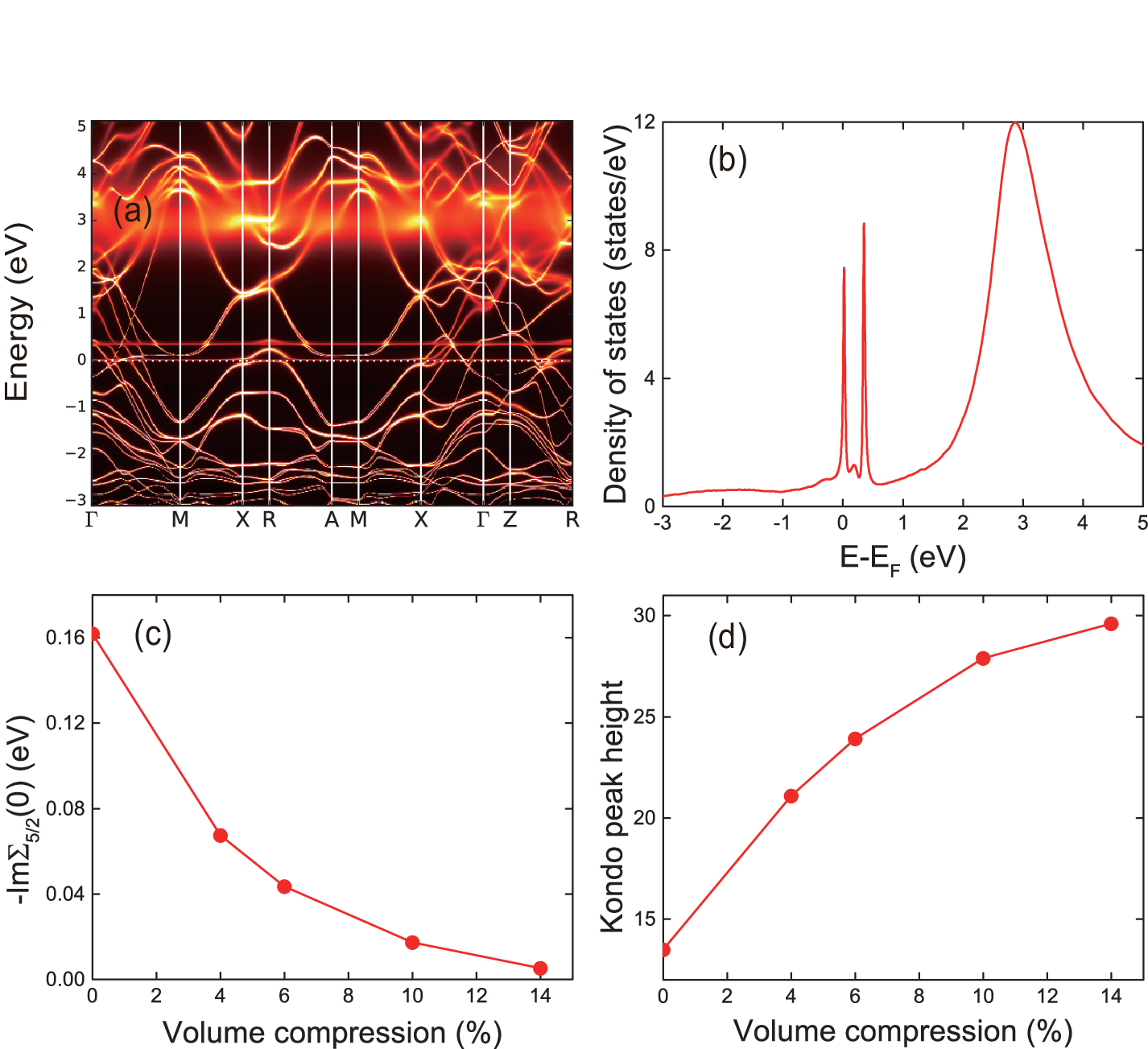}
\caption{
(a) and (b) show the momentum-resolved spectral function and Ce-4$f$ density of states of CeRh$_2$As$_2$ at 20 K under ambient pressure, respectively, in which the Kondo resonance peak, the lower and upper Hubbard bands can be clearly identified. (c) and (d) illustrate the zero-frequency self-energy and height of Kondo peak as functions of compression ratio at 20 K, respectively.}
\label{add1}
\end{figure*}

Firstly, we discuss CeRh$_2$As$_2$. Fig. \ref{SigmaDelta}(a) shows the hybridization function $\Delta(\omega)$ on real-frequency axis at 20 K, which is related to the imaginary part of impurity hybridization function $\tilde{\Delta}(\omega)$ by $\Delta(\omega)=-\frac{1}{\pi}\mathrm{Im}\tilde{\Delta}(\omega)$. It can be seen that as the compression rate increases, the hybridization function exhibits significant enhancement in wide energy range, indicating that the $c$-$f$ Kondo hybridization strength also increases accordingly. As shown in Fig. \ref{SigmaDelta}(b), at 20 K, the imaginary part of Ce-4$f_{5/2}$ self-energy $-\mathrm{Im}\Sigma_{5/2}(\omega)$ exhibits an evident dip structure around zero frequency, which induces sharp Kondo resonance peak near the Fermi level at low temperature, as shown in Fig. \ref{DOS}(a) and \ref{DOS}(b), in which the additional peaks around 0.36 eV above $E_F$ are contributed by 4$f_{7/2}$ state. As the compression rate increases, the zero-frequency magnitude of $-\mathrm{Im}\Sigma_{5/2}(\omega)$ decreases rapidly (reaches 5.2 meV at 14$\%$ volume compression), leading to significant enhancement of Kondo resonance peak, as illustrated in Fig. \ref{DOS}(c). Above 6$\%$ compression rate, Ce-4$f_{5/2}$ self-energy can be well fitted by a parabolic function near zero frequency as $-\mathrm{Im}\Sigma_{5/2}(\omega)\approx \alpha(\omega-\omega_0)^2+\Sigma_0$ with very small $\omega_0$ and $\Sigma_0$, signaling the appearance of heavy Fermi liquid behavior as a result of enhanced Kondo hybridization under compression~\cite{Ma23}.

\begin{figure*}[tbp]
\hspace{-0cm} \includegraphics[totalheight=5in]{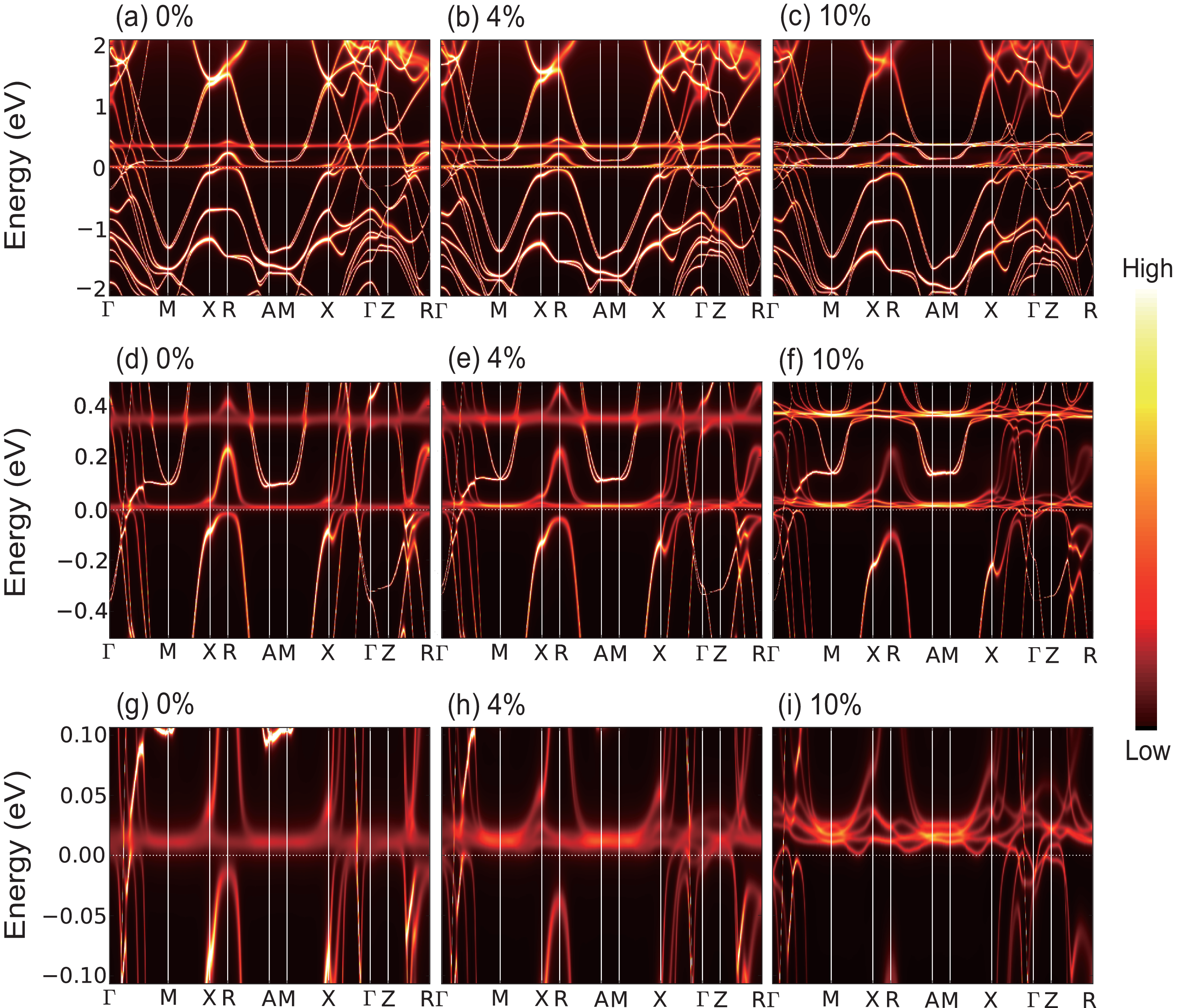}
\caption{
DFT+DMFT momentum-resolved spectral function of CeRh$_2$As$_2$ at 20 K, under 0$\%$, 4$\%$ and 10$\%$ volume compressions, respectively. From top to bottom rows, the energy windows are zoomed-in close to the Fermi level. The gradually distinguishable heavy-fermion bands near the Fermi level verify a localized-to-coherent crossover of Ce-4$f$ electrons under compression.
}
\label{specfunc_20K}
\end{figure*}

\begin{figure*}[tbp]
\hspace{-0cm} \includegraphics[totalheight=4.0in]{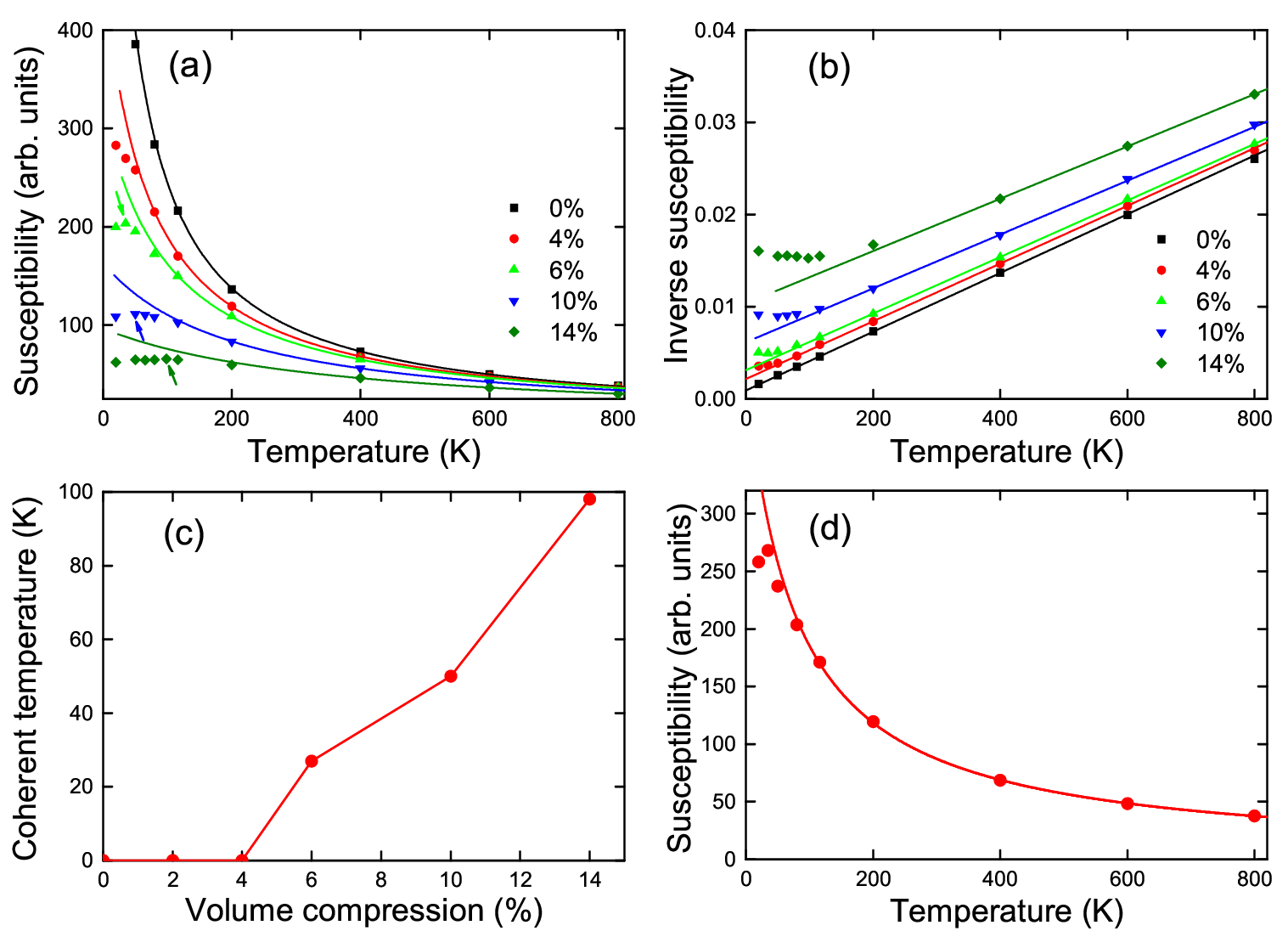}
\caption{
Solid dots denote DFT+DMFT temperature dependence of (a) local spin susceptibility $\chi_s$ and (b) inverse susceptibility $\chi^{-1}_s$ of CeRh$_2$As$_2$ at various volume compression rates.
In (a), the arrows marking the maximum of $\chi_s$ indicate the onset of coherent temperatures, which first appears at 6$\%$ volume compression. (c) Coherent temperature $T_{\mathrm{coh}}$ vs volume compression percentage. (d) Temperature dependence of local spin susceptibility of CeIr$_2$As$_2$ under ambient pressure through DFT+DMFT calculation, suggesting its coherent temperature $T_{\mathrm{coh}}\approx35 $ K. In (a), (b) and (d), the solid lines are the fitted Curis-Weiss functions.
}
\label{Sus_Tcoh}
\end{figure*}

The appearance of Kondo resonance can be witnessed via momentum-resolved spectral function $A(\mathbf{k},\omega)$, which can be related directly to ARPES measurements. Fig. \ref{add1} (a) illustrates $A(\mathbf{k},\omega)$ between [-3, 5] eV, in which the broad lower and upper Hubbard bands appear around -2 and 3 eV, respectively, roughly correspond to $\pm U/2$. From Fig. \ref{add1} (b), the spectral weight of the lower Hubbard band is quite weak, while the weight of the upper Hubbard band is rather intense, in accord with the characteristic of many Ce-based materials.
Fig. \ref{specfunc_20K} shows $A(\mathbf{k},\omega)$ of CeRh$_2$As$_2$ at volume-compression ratios from 0$\%$, 4$\%$ to 10$\%$. The hybridization of conduction electrons with 4$f_{5/2}$ or 4$f_{7/2}$ states give rise to two groups of flat heavy-fermion bands near the Fermi level (set as 0 eV) and 0.36 eV above, respectively. Since $c$-$f$ hybridization is enhanced with increasing compression, the spectral weight of these heavy-fermion bands become gradually intense and are eventually clearly resolved at 10$\%$ compression rate, see the enlarged view in Fig. \ref{specfunc_20K}(i). In addition, the nonsymmorphic symmetries in the space group of CeRh$_2$As$_2$ (P4/$nmm$, No. 129) preserve Dirac-type band crossings along X-R and M-A paths in the Brillouin zone, which can be clear seen in Fig. \ref{specfunc_20K}(i). Therefore, under compression, CeRh$_2$As$_2$ can be classified as a heavy node-line Dirac semimetal, similar to the case of CePt$_2$Si$_2$~\cite{Ma23}.

The formation of sharply resolved heavy-fermion bands at 10$\%$ volume compression (Fig. \ref{specfunc_20K}(i)) indicates that now the local moments are well screened by conduction electrons and Ce-4$f$ electrons become fully itinerant, i.e, a localized-itinerant crossover takes place~\cite{Lu16}, leading to heavy Fermi liquid state. To evaluate the compression rate and temperature at which the localized-itinerant crossover occurs, we calculate the local spin susceptibility $\chi_s$ of Ce-4$f$ states during DFT+DMFT iterations, and the results are displayed in Fig. \ref{Sus_Tcoh}. Fig. \ref{Sus_Tcoh}(a) and \ref{Sus_Tcoh}(b) show the temperature dependence of $\chi_s$ and its inverse $\chi^{-1}_s$, respectively, at five different compression rates ranging from 0$\%$ to 14$\%$, in which the solid lines denote the fitted Curis-Weiss functions $\chi_s=C/(T+\theta)$. The coincidence of $\chi_s$ dots with Curis-Weiss line at 0$\%$ compression confirms the local-moment nature of CeRh$_2$As$_2$ at ambient pressure.
At low compression rate (4$\%$), the susceptibility only slightly diverges from Curis-Weiss line at low temperature, while above 6$\%$ compression rate, a local maximum (at 27 K) starts to arise on $\chi_s$-$T$ curve, illustrating the onset of coherent temperature $T_{\mathrm{coh}}$ at which the localized-itinerant crossover occurs upon cooling~\cite{Nam21,Ma23}. In Fig. \ref{add1} (c-d), the sharp decrease of zero-frequency self-energy and significant enhancement of Kondo-peak height between 4$\%$ to 6$\%$ volume compression further confirm the crossover to Kondo coherence.
In Fig. \ref{Sus_Tcoh}(c), $T_{\mathrm{coh}}$ is plotted with varying compression ratios, which clear indicates that the localized-itinerant crossover states to emerge around 6$\%$ volume compression.

\begin{figure*}[tbp]
\hspace{-0cm} \includegraphics[totalheight=1.5in]{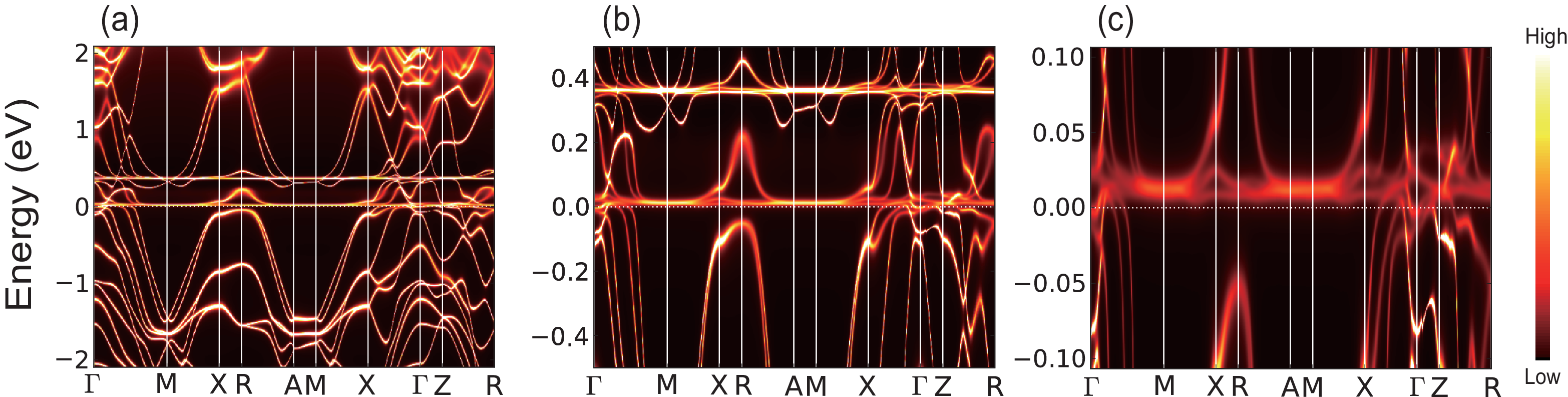}
\caption{
DFT+DMFT momentum-resolved spectral function of CeIr$_2$As$_2$ at 20 K under ambient pressure (displayed in zoomed-in energy range near the Fermi level from (a) to (c)). Heavy-fermion bands are already clearly resolved even at ambient pressure.
}
\label{CeIr2As2}
\end{figure*}

In order to analyse the consequence of localized-itinerant crossover to the heavy-fermion SC state, we also calculate the local spin susceptibility $\chi_s$ of non-superconducting CeIr$_2$As$_2$~\cite{Pfannenschmidt12,Cheng19}, which is isomorphic to CeRh$_2$As$_2$. As shown in Fig. \ref{Sus_Tcoh}(d), under ambient pressure, the temperature dependence of $\chi_s$ for CeIr$_2$As$_2$ already possesses a maximum, suggesting the itinerant nature of the Ce-4$f$ states below 35 K, which can be intuitively illustrated by the sharply dispersive heavy-fermion bands in its spectral function given in Fig. \ref{CeIr2As2}. The full itineration of $f$ electrons leads to heavy Fermi liquid state, which arises at higher pressure than the SC state in pressure-temperature phase diagram of heavy-fermion SC compounds~\cite{Paschen21,Nica22}.
Therefore, one can speculate that the itineration of $f$ electrons may take disadvantage of the formation of heavy-fermion superconductivity, as will be discussed in detail in the following section.

\begin{figure*}[tbp]
\hspace{-0cm} \includegraphics[totalheight=4.1in]{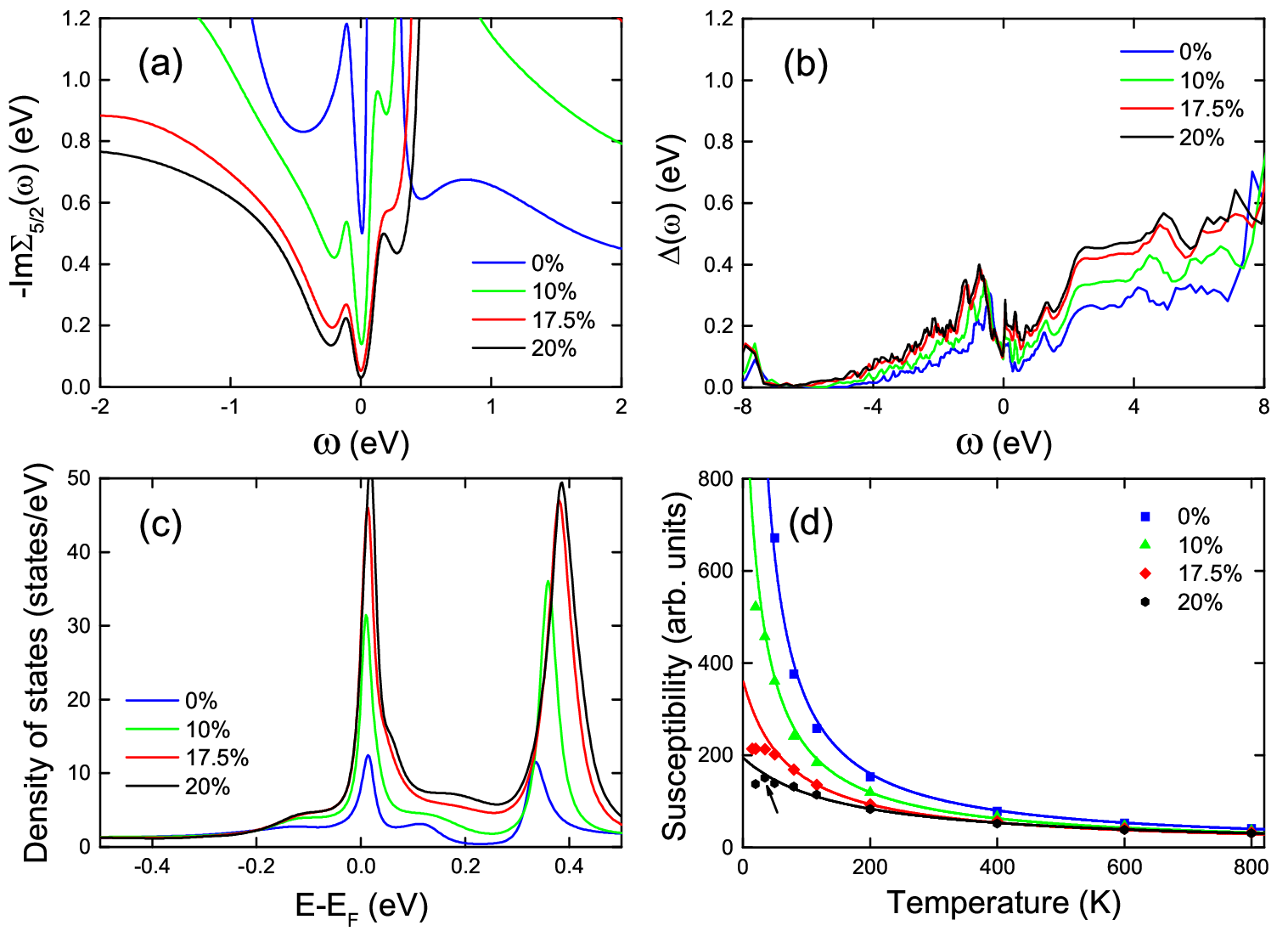}
\caption{
(a) and (b) show the real-frequency imaginary self-energy $-\mathrm{Im}\Sigma_{5/2}(\omega)$ and hybridization function $\Delta(\omega)$ of CeSb$_2$, respectively, at four compression percentages ranging from 0$\%$ to 20$\%$, at 20 K. (c) As the compression rate increases, the Kondo resonance peak near the Fermi level (at 20 K) is greatly enhanced. (d) Temperature dependence of local spin susceptibility $\chi_s$ at four volume-compression ratios. The solid dots denote DFT+DMFT results, and the solid lines are the fitted Curis-Weiss functions.
In (d), the black arrow marking the maximum of $\chi_s$ indicates a coherent temperatures $T_{\mathrm{coh}}\approx35$ K at 20$\%$ volume compression percentage.
}
\label{CeSb2}
\end{figure*}

\begin{figure*}[tbp]
\hspace{-0cm} \includegraphics[totalheight=5.0in]{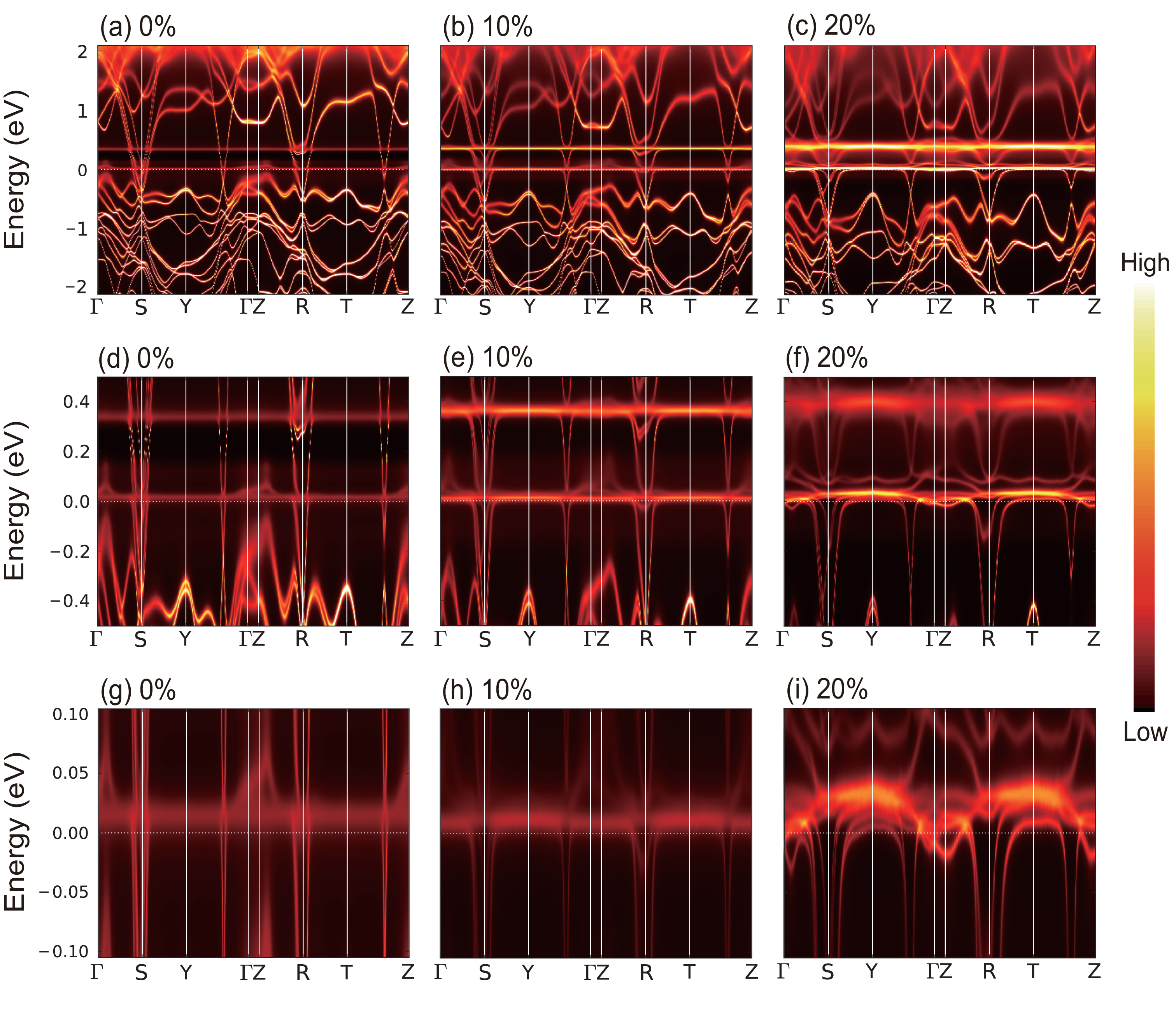}
\caption{
Momentum-resolved spectral function of CeSb$_2$ at 20 K, under 0$\%$, 10$\%$ and 20$\%$ volume-compression ratios, respectively. From top to bottom rows, the energy ranges are zoomed-in close to the Fermi level for clearer view. The heavy-fermion bands are well distinguishable near the Fermi level at 20$\%$ compression rate.
}
\label{CeSb2_specfunc}
\end{figure*}

Now we turn to another recently discovered heavy-fermion SC material CeSb$_2$~\cite{Squire22}. The DFT+DMFT results are displayed in Figs. \ref{CeSb2} and \ref{CeSb2_specfunc}. In comparison with the case of CeRh$_2$As$_2$, volume compression causes similar impacts to CeSb$_2$, i.e., it reduces the imaginary Ce-4$f_{5/2}$ self-energy $-\mathrm{Im}\Sigma_{5/2}(\omega)$ (see Fig. \ref{CeSb2}(a)), enhances the Kondo resonance peak (Fig. \ref{CeSb2}(c)) and hybridization strength (see Fig. \ref{CeSb2}(b)) remarkably, hence eventually induces a localized-itinerant crossover at about 35 K at 20$\%$ compression rate (see the black arrow indicating the maximum of $\chi_s$ in Fig. \ref{CeSb2}(d)). Consequently, above the critical volume compression ratio 20$\%$, well-defined hybridization bands show up in the momentum-resolved spectral function near the Fermi level below the coherent temperature $T_{\mathrm{coh}}=35$ K (see Fig. \ref{CeSb2_specfunc}(i)).
In contrast to CeRh$_2$As$_2$, the critical compression rate of CeSb$_2$ which starts to induce localized-itinerant crossover at non-zero coherent temperature is much higher, which can be ascribed to stronger local-moment character of Ce-4$f$ electrons in CeSb$_2$ than in CeRh$_2$As$_2$ at ambient pressure, since at 0$\%$ volume compression, the 4$f$ resonance peak contributed from one Ce atom in CeSb$_2$ is much lower than in CeRh$_2$As$_2$ (compare Fig. \ref{DOS}(a) with Fig. \ref{CeSb2}(c) and note that there are much more Ce atoms in the unit cell of CeSb$_2$ than in CeRh$_2$As$_2$). Different degrees of 4$f$ localization can give rise to distinct ground state phases of CeRh$_2$As$_2$ and CeSb$_2$ at ambient pressure, as will be discussed below. In above calculations, the crystal-field splitting (CFS) of $f$ orbits has been pre-examined and found to be one more orders of magnitude
smaller than SOC splitting (several meV vs 0.36 eV) and is compatible with Kondo coherent scale. The Kondo coherent temperature $T_{\mathrm{coh}}$ is sensitive to the effective degeneracy of the Kondo problem, which is affected by CFS, thus by including CFS in DFT+DMFT calculations, the critical volume compression ratios in CeRh$_2$As$_2$ and CeSb$_2$ will be shifted, which is expected to give only quantitative affect to the phase evolution, thus, for simplicity, CFS is not considered explicitly at present work~\cite{Nam19,Zhu20,Lu16}, and for further studies, CFS should be included properly in DFT+DMFT calculation~\cite{Jang22}.

\begin{figure*}[tbp]
\hspace{-0cm} \includegraphics[totalheight=3.9in]{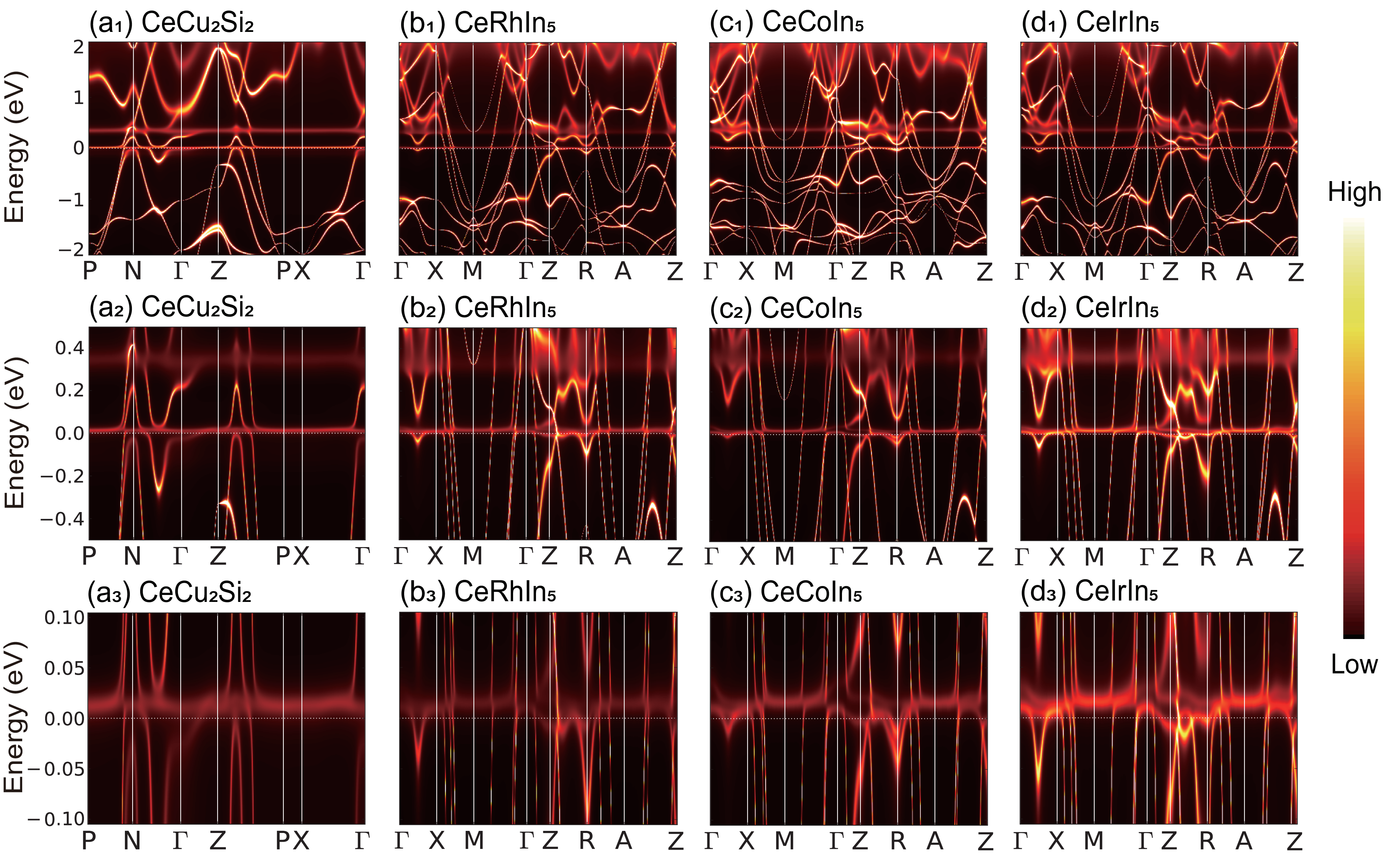}
\caption{
Momentum-resolved spectral functions of CeCu$_2$Si$_2$ and CeMIn$_5$ (M=Rh, Co, Ir) at 20 K. From top to bottom rows, the energy windows are zoomed-in close to the Fermi level. The hybridization bands of CeRhIn$_5$ are quite blurred, while for CeCoIn$_5$ and CeIrIn$_5$~\cite{Chen17,Chen18}, they become much more intense than CeRhIn$_5$. The spectral weights of the hybridization bands can be roughly reflected by Ce-$f$ density of states in Fig. \ref{dlt_phase}(a).
}
\label{CeCu2Si2_CeCoIn5}
\end{figure*}

\begin{figure}[tbp]
\hspace{0cm} \includegraphics[totalheight=2.1in]{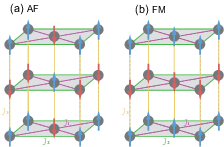}
\caption{
DFT+U magnetic configurations for calculating RKKY strengths in quasi-two dimensional CeMIn$_5$ (M= Rh, Co, Ir). The solid dots represent Ce atoms with local spins. (a) and (b) are intra-plane AF and FM states, respectively. The nearest-neighbor exchange $J_1$ corresponds to RKKY interaction $J_H$ to be evaluated.
}
\label{RKKY}
\end{figure}

\begin{figure*}[tbp]
\hspace{0cm} \includegraphics[totalheight=4.2in]{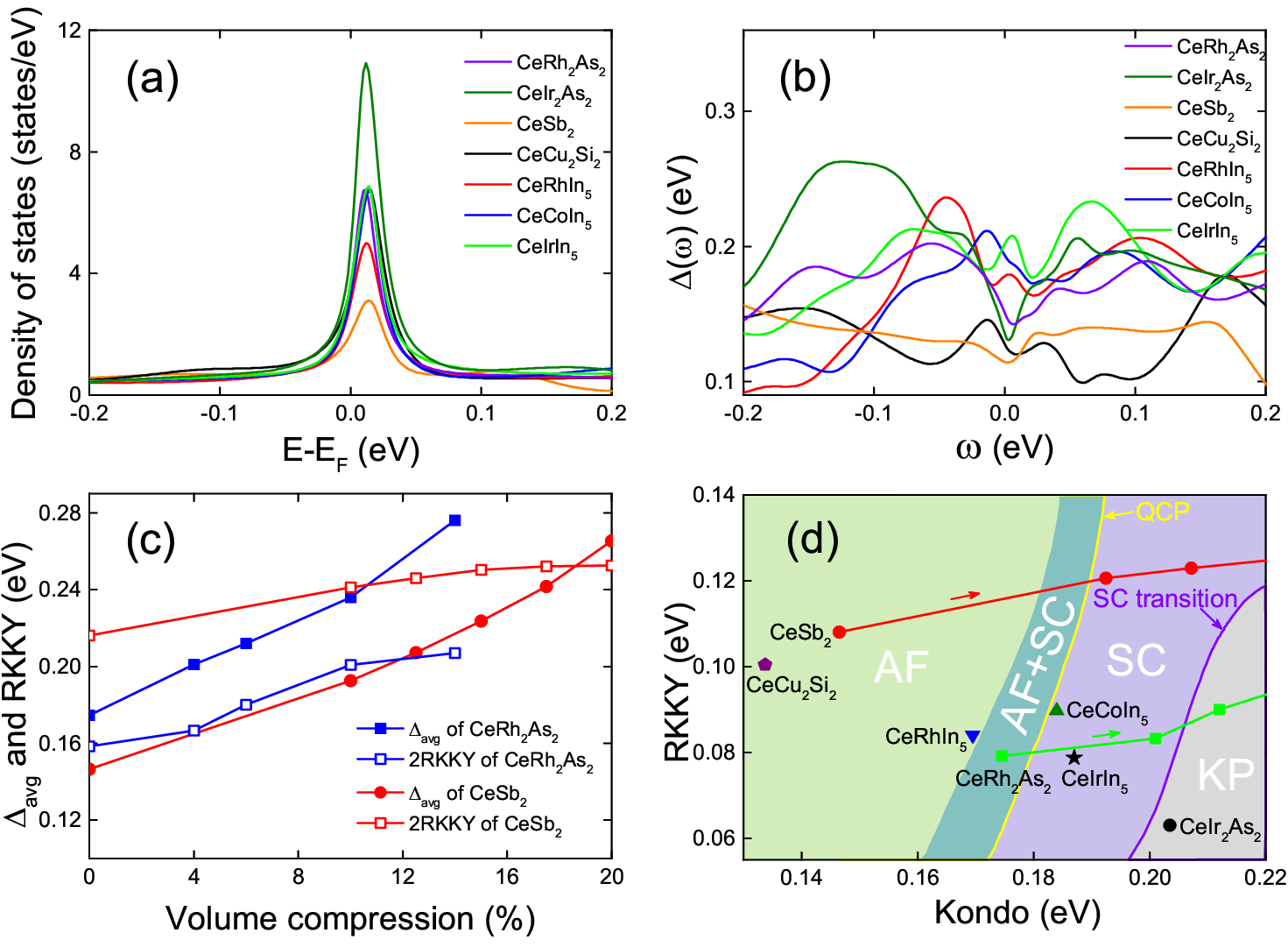}
\caption{
(a) and (b) show the 4$f$ density of states $\rho(\omega)$ per Ce atom and the hybridization function $\Delta(\omega)$ at 20 K under ambient pressure, respectively. (c) Average hybridization $\Delta_{\mathrm{avg}}$ and RKKY strength $J_H$ of CeRh$_2$As$_2$ and CeSb$_2$ vs volume compression, at 20 K. (d) Kondo coupling $J_K$ and RKKY strength $J_H$ in various compounds at 20 K, while the variations under compression for CeRh$_2$As$_2$ and CeSb$_2$ are indicated by green and red arrows, respectively. Phases and the evolution under pressure are determined through competition of $J_K$ and $J_H$, see Tab. \ref{tab1} and text.
}
\label{dlt_phase}
\end{figure*}

\begin{figure*}[tbp]
\hspace{-2.5cm} \includegraphics[totalheight=3.2in]{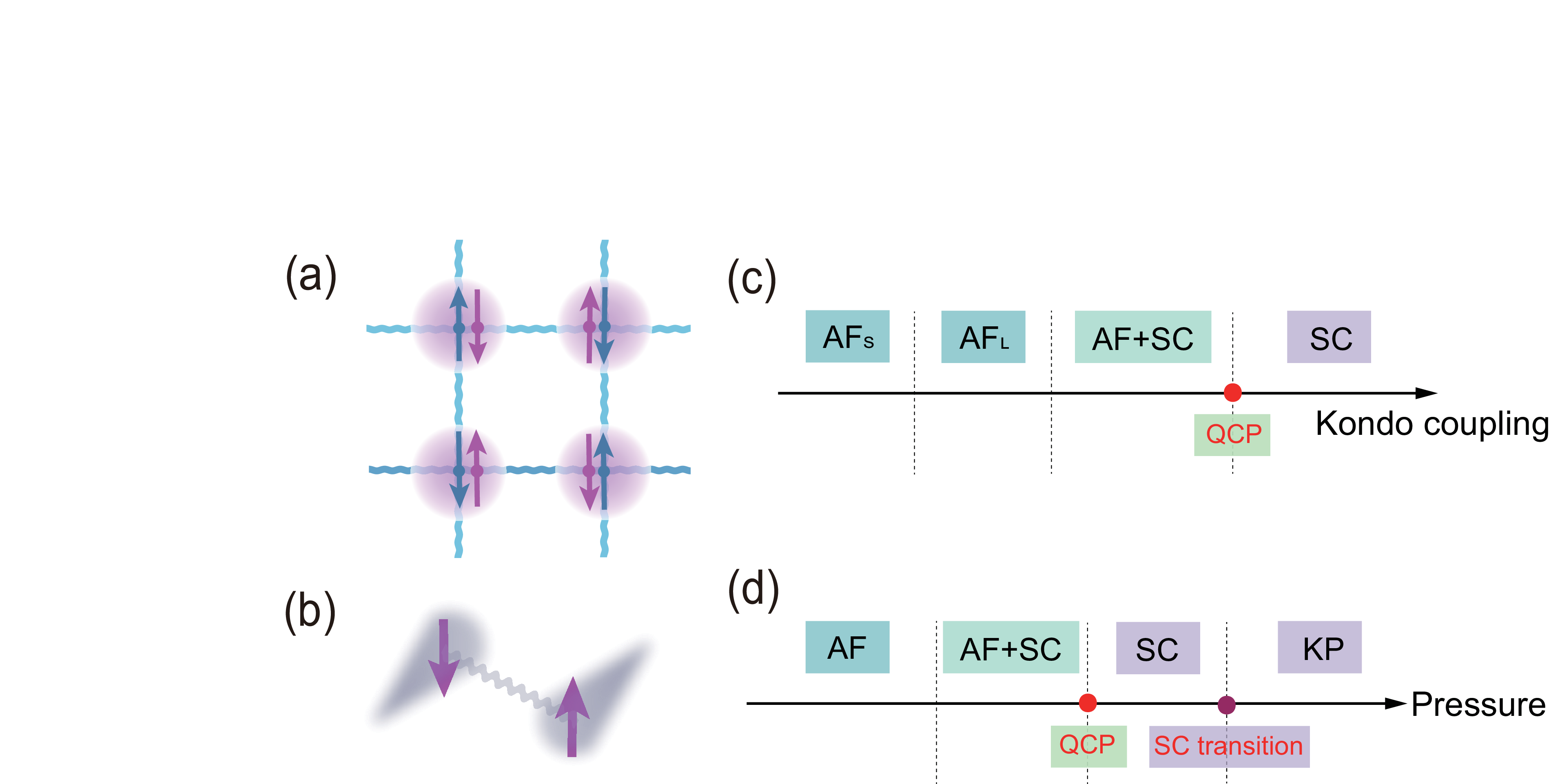}
\caption{
Schematic description of heavy-fermion superconductor. (a) The interplay of short-range singlet pairing between $f$ electrons (blue arrows) and Kondo screening by conduction electrons (purple arrows) drives (b) Cooper pairing between heavy quasi-particles. (c) Phase evolution in the effective model with increasing Kondo coupling. The magnetic QCP separates the SC phases into AF+SC coexisting phase and paramagnetic SC phase. (d) Phase evolution with pressure in Ce-based superconductors. Beside the magnetic QCP, a SC transition from paramagnetic SC phase to KP phase driven by localized-itinerant crossover takes place at higher pressure. In (d), AF$_s$ and AF$_L$ states are uniformly denoted by AF state.
}
\label{ModelDescription}
\end{figure*}

\section{effective-model description of phase evolution under pressure}
In order to give a concise description concerning the general feature of the phase evolution of heavy-fermion superconductors under pressure, and to understand the consequence of the pressure-induced localized-itinerant crossover to the SC state as well, we construct a minimal effective model for heavy-fermion SC systems. Since in the heavy-fermion SC compounds we considered, Ce-4$f$ electrons are mainly localized under ambient pressure, for simplicity, the role of Ce-4$f$ electrons are treated as local moments as a starting point. The model Hamiltonian reads
\begin{equation}
\mathcal{H}=\sum_{\mathbf{k},\sigma }\epsilon _{\mathbf{k}}c_{\mathbf{
k}\sigma }^{\dag }c_{\mathbf{k}\sigma }+J_K\sum_{i}\mathbf{S}_{i}\cdot \mathbf{
S}_{ic}+J_H\sum_{\langle ij\rangle}\mathbf{S}_{i}\cdot \mathbf{
S}_{j},\label{H}
\end{equation}
 which includes two dominating interactions in heavy-fermion systems, namely the Kondo coupling between conducting $c$ electrons and local moments with strength $J_K$, and the nearest-neighbor RKKY superexchange $J_{H}$ term between local moments.
 In fact, similar simplified models have been extensively adopted in the literature, and have revealed rich common features of phase transitions in heavy-fermion systems~\cite{Doniach77,Si10}, although more precise multi-orbital models should be fitted to explain experimental behaviors of individual materials. As many typical heavy-fermion SC compounds such as CeMIn$_5$ (M=Co, Rh) (Fig. \ref{lattice}(d)) exhibit quasi-two-dimensional Fermi surfaces~\cite{Cornelius00,Koitzsch08}, we consider above model on a two-dimensional square lattice, which can represent the arrangement of Ce atoms on $\mathbf{a}$$\mathbf{b}$ plane (see Fig. \ref{lattice}), with tight-binding dispersion of conduction electrons
$
\epsilon _{\mathbf{k}}=-2t(\cos {k_{x}}+\cos {k_{y}})+4t^{\prime
}\cos {k_{x}}\cos {k_{y}}-\mu,
$
in which $t$ and $t^\prime$ are the nearest-neighbor and next-nearest-neighbor hopping strengths, respectively, and $\mu$ is the chemical potential for fixing conduction-electron number $n_c$.

In the context of $c$-$f$ hybridization, the spin density of local moments can be expressed via slave-fermion representation as $\mathbf{S}_{i}=\frac{1}{2}\sum_{\alpha\beta}f^\dag_{i\alpha}\mathbf{\sigma}_{\alpha\beta}f_{i\beta}$ with $\mathbf{\sigma}$ being the Pauli matrix, and $f$ operators are subjected to the constraint $\sum_{\sigma}f^{\dag}_{i\sigma}f_{i\sigma}=1$, imposing by adding a Lagrangian term $\sum_i\lambda_i(\sum_{\sigma}f^{\dag}_{i\sigma}f_{i\sigma}-1)$ to Eq. \ref{H}. In such representation, the RKKY interaction can be rewritten equivalently as
\begin{eqnarray}
\mathbf{S}_{i}\cdot \mathbf{S}_{j}&=&-\frac{1}{2}
(f_{i\uparrow }^{\dag }f_{j\downarrow }^{\dag}-f_{i\downarrow }^{\dag
}f_{j\uparrow }^{\dag})(f_{j\downarrow }f_{i\uparrow }-f_{j\uparrow
}f_{i\downarrow })+\frac{1}{4},
\end{eqnarray}
which represents singlet-pairing interacting between neighboring $f$ electrons, and can be decoupled by introducing the singlet-pairing strength
\begin{align}
\Delta=\langle f_{i \uparrow }^{\dag}f_{i\pm \mathbf{x} \downarrow}^{\dag}-
            f_{i \downarrow }^{\dag}f_{i\pm \mathbf{x} \uparrow}^{\dag}\rangle=-\langle f_{i \uparrow }^{\dag}f_{i\pm \mathbf{y} \downarrow}^{\dag}-
            f_{i \downarrow }^{\dag}f_{i\pm \mathbf{y} \uparrow}^{\dag}\rangle,
\end{align}
which features $d_{x^2-y^2}$ symmetry, as many typical heavy-fermion SC systems carry $d$-wave symmetry in there SC pairing~\cite{An10,Akbari12,Zhou13,Allan13}. $\Delta$ is a typical form of short-range magnetic correlation, and in previous studies, it has been verified that similar magnetic correlations could induce heavy-fermion superconductivity and other unconventional superconductivity, such as SC states in cuprates and nickelates~\cite{Liu12,Yang23}. For heavy-fermion systems, at presence of Kondo hybridization, the magnetic correlation can drive a paramagnetic heavy-fermion SC state~\cite{Liu12,Dyke14}, in which the SC pairing function $\langle
c^{\dag}_{\mathbf{k}\uparrow}c^{\dag}_{-\mathbf{k}\downarrow}
+c_{-\mathbf{k}\downarrow}c_{\mathbf{k}\uparrow}
\rangle$ features $d$-wave symmetry, and the SC order parameter can be evaluated as
\begin{align}
\Delta_{sc}&=\frac{1}{2N}\sum_{\mathbf{k}}(\cos \mathbf{k}_x-\cos \mathbf{k}_y)
\langle
c^{\dag}_{\mathbf{k}\uparrow}c^{\dag}_{-\mathbf{k}\downarrow}
+c_{-\mathbf{k}\downarrow}c_{\mathbf{k}\uparrow}
\rangle.
\label{SC}\end{align}
In the following, we will show that magnetic correlation can also drive heavy-fermion SC state at presence of AF long-range magnetic order, leading to an AF+SC coexisting phase in certain parameter region. Two AF order parameters can be introduced as
\begin{align}
 &m_c=-\frac{1}{2}\langle\sum_{\sigma}\sigma c^{\dag}_{i\sigma}c_{i\sigma}\rangle e^{i\mathbf{Q}\cdot\mathbf{R}_i},\nonumber\\
  &m_f=\frac{1}{2}\langle\sum_{\sigma}\sigma f^{\dag}_{i\sigma}f_{i\sigma}\rangle e^{i\mathbf{Q}\cdot\mathbf{R}_i},
 \label{mcmf}\end{align}
 where $\mathbf{Q}=(\pi,\pi)$ is the AF vector, then the staggered magnetization in the AF state can be expressed by $M=m_f-m_c$. The Kondo coupling term can be decomposed as~\cite{Li15}
\begin{align}
\mathbf{S}_{i}\cdot \mathbf{S}_{ic}=& -\frac{3}{8}
(c_{i\uparrow }^{\dag }f_{i\uparrow }+c_{i\downarrow }^{\dag
}f_{i\downarrow })(f_{i\uparrow }^{\dag }c_{i\uparrow }+f_{i\downarrow
}^{\dag }c_{i\downarrow })  \nonumber\\
& +\frac{1}{8}(c_{i\uparrow }^{\dag }f_{i\uparrow }-c_{i\downarrow
}^{\dag }f_{i\downarrow })(f_{i\uparrow }^{\dag }c_{i\uparrow
}-f_{i\downarrow }^{\dag }c_{i\downarrow })  \nonumber\\
& +\frac{1}{8}(c_{i\uparrow }^{\dag }f_{i\downarrow }+c_{i\downarrow
}^{\dag }f_{i\uparrow })(f_{i\downarrow }^{\dag }c_{i\uparrow }+f_{i\uparrow
}^{\dag }c_{i\downarrow }) \nonumber\\
& +\frac{1}{8}(c_{i\uparrow }^{\dag }f_{i\downarrow }-c_{i\downarrow
}^{\dag }f_{i\uparrow })(f_{i\downarrow }^{\dag }c_{i\uparrow }-f_{i\uparrow
}^{\dag }c_{i\downarrow }),\label{J}
\end{align}
in which the first term represents local singlet $c$-$f$ hybridization, while the last three terms denote triplet hybridizations. At presence of AF order, the dominating $c$-$f$ hybridization strength $\langle c^{\dag}_{i\sigma}f_{i\sigma}\rangle$ can vary with sublattices and spin directions, leading to two different hybridization parameters
$V_{s}=\frac{1}{2}\sum_{\sigma}\langle c^\dag_{i\sigma} f_{i\sigma}\rangle$ and
$V_{t}
=\frac{1}{2}\sum_{\sigma}\sigma e^{i\mathbf{Q}\cdot\mathbf{R}_i} \langle c^\dag_{i\sigma} f_{i\sigma}\rangle$, where $V_s$ is uniform singlet hybridization and $V_t$ is staggered triplet hybridization, and the latter requires breaking of particle-hole symmetry of conduction electrons at presence of AF order. Then the Kondo coupling in Eq. \ref{J} can be treated by performing Hartree-Fock approximation using $V_s$ and $V_t$. In addition, it has been shown from Monte Carlo simulations of Kondo lattice model and periodic Anderson model that Kondo coupling can naturally induce AF order~\cite{Vekic95,Assaad99}. In the mean-field treatment, such AF order can be induced by decoupling the polarization term in Kondo interaction ($J_{K}\mathbf{S}^z_i\mathbf{S}^z_{ic}$) by substituting Eq. \ref{mcmf}, to be
\begin{align}
\sum_{i}\mathbf{S}^z_{i}\cdot \mathbf{S}^z_{ic}=
&-\frac{1}{2}m_c\sum_{\mathbf{k}\sigma}\sigma f_{\mathbf{k}+\mathbf{Q}\sigma}^{\dag}f_{\mathbf{k}\sigma}\nonumber\\&+
\frac{1}{2}m_f{\sum_{\mathbf{k} \sigma}}\sigma{c_{\mathbf{k}+\mathbf{Q}\sigma}^{\dag}}c_{\mathbf{k}\sigma}
+m_cm_f,
\end{align}
which can lead to proper description of magnetization~\cite{Li15,Li16}. For convenient, the subsequent derivation of self-consistent equations which determine the chemical potential $\mu$, Lagrange multiplier $\lambda$, and mean-field parameters $\Delta$, $V_s$, $V_t$, $m_c$, $m_f$, and the discussion of numerical results are given in detail in the appendix.

\heavyrulewidth=1bp

\begin{table*}
\small
\renewcommand\arraystretch{1.3}
\caption{\label{tab1}
Kondo coupling (evaluated through average hybridization $\Delta_{\mathrm{avg}}$) and RKKY interaction via first-principle calculations. The predictions for ground-state phases are discussed in the text, and are compared with experimental observations.}
\begin{tabular*}{17cm}{@{\extracolsep{\fill}}cccccccccc}
\toprule
        & CeRh$_2$As$_2$  &  CeSb$_2$ & CeIr$_2$As$_2$ & CeCu$_2$Si$_2$  & CeRhIn$_5$ &CeCoIn$_5$ &CeIrIn$_5$  \\
\hline
 Kondo (eV) & 0.1745 &  0.147  &0.2035  & 0.1338 & 0.1695  &0.184  &0.187  \\
     & medium & weak   &  strong   & weak     & medium-low     &   strong    &  strong\\
 RKKY (eV)       & 0.0792  & 0.1081 & 0.0631    & 0.1004     & 0.0841    &0.0897    &0.0788  \\
 & medium  & strong  &  weak   & strong     & medium     &   medium    &  medium\\
 Kondo/2RKKY  &  1.102 &  0.681  &  1.613  &  0.666   &  1.01  &  1.03   & 1.19\\
 Prediction   & AF+SC  & AF    &  KP     & AF       & AF     &SC       &SC  \\
       &     &   & $T_{\mathrm{coh}}=35\mathrm{K}$     &      &     &    &  \\
  Experiment   & AF+SC~\cite{Siddiquee22,Kibune22}  & AF~\cite{Squire22} & KP~\cite{Cheng19}  & AF~\cite{Yuan03} & AF~\cite{Ida08}  &SC~\cite{Sidorov02}  &SC~\cite{Shang14}  \\
    & &    & $T_{\mathrm{coh}}=65\mathrm{K}$  &  &  & &  \\
\bottomrule
\end{tabular*}
\label{tab1}
\end{table*}
Based on the results of model calculations, we can draw a schematic description of heavy-fermion SC systems in Fig. \ref{ModelDescription}. We find that in addition to the Kondo hybridization ($V_s$ and $V_t\neq0$) between $c$ and $f$ electrons, the RKKY superexchange can lead to nearest-neighbor singlet pairing ($\Delta\neq0$) between $f$ electrons (see Fig. \ref{ModelDescription}(a)), which combines with $c$-$f$ hybridization to produce Cooper pairing ($\Delta_{sc}\neq0$) between dressed heavy quasiparticles (see Fig. \ref{ModelDescription}(b)), thus inducing the heavy-fermion SC state. Remarkably, even at presence of AF long-range magnetic order, the short-range singlet pairing can persist in some parameter region, resulting in an AF+SC coexisting phase with $\Delta_{sc}$ and $M\neq0$. We have calculated the evolution of ground-state phases with Kondo coupling $J_K$ (see appendix), the numerical results show that at weak Kondo coupling, long-range AF order dominates ($M\neq0$), and the $f$ electrons are fully localized and decoupled with $c$ electrons ($V_s=0$), leading to AF$_s$ phase with small Fermi surface constructed only by $c$ electrons. While $J_K$ increases, the $c$-$f$ hybridization sets in ($V_s\neq0$) with AF order, leading to AF$_L$ phase with large Fermi surface forming by hybridized heavy fermions. While $J_K$ is further enhanced, the long-range AF order is suppressed partially, then the arising of short-range singlet pairing of $f$ electrons ($\Delta\neq0$) can gain lower energy than the AF$_L$ state, and by combining with Kondo hybridization, drives an AF+SC coexisting phase with $\Delta_{sc}, M\neq0$. At a critical Kondo coupling strength $J^{c1}_{K}$, the AF order is fully suppressed, leading to an AF magnetic transition, in the meanwhile, $\Delta$ survives at and above $J^{c1}_{K}$, thus produces a paramagnetic heavy-fermion SC state beyond the magnetic transition point.
Therefore, the magnetic transition at $J^{c1}_{K}$ can be related to the magnetic QCP in heavy-fermion SC compounds around which the SC phase emerges~\cite{Millis93,Arndt11,Park06,Squire22}, see Fig. \ref{ModelDescription}(c). In the SC phase, increase of $J_K$ will reduce the magnetic correlation $\Delta$ and give rise to fast reduction of SC order, see Fig. \ref{ModelCalculation}(a).

To track the ground-state phase evolution of realistic heavy-fermion superconductors under pressure, we combine the results of effective model with the first-principle results. In above section, we have demonstrated that the volume compression can remarkably enhance the hybridization function $\Delta(\omega)$ in wide energy range (see Figs. \ref{SigmaDelta}(a) and \ref{CeSb2}(b)).
Fig. \ref{dlt_phase}(c) illustrates the averaged hybridization strength, which shows a notable enhancement with compression ratio. Since in Anderson impurity model, the hybridization strength $\Delta_{\mathrm{avg}}$ and Kondo coupling $J_\mathrm{K}$ are both square proportional to $c$-$f$ hopping $V_{cf}$ as $\Delta_{\mathrm{avg}}, J_\mathrm{K}\sim V^2_{cf}$ (by Schrieffer-Wolff transformation), $\Delta_{\mathrm{avg}}$ can be closely related to Kondo coupling strength, thus indicating a remarkable increase of $J_\mathrm{K}$ with pressure. The hybridization functions of various materials are given in Fig. \ref{dlt_phase} (b), with corresponding $\Delta_{\mathrm{avg}}$ plotted in Fig. \ref{dlt_phase}(d), and are also explicitly listed in Tab. \ref{tab1}, which reflect their Kondo coupling strength.

In order to evaluate the strength of RKKY exchange $J_H$, we perform the DFT+U calculations for the magnetic phases. As can be seen from the lattice structures in Fig. \ref{lattice}, the Ce atoms in these SC compounds are arranged into quasi-two-dimensional lattices, with intra-plane square lattice structures (slightly distorted in CeSb$_2$). Therefore, the largest exchange coupling between Ce atoms follows the $\mathbf{a}$ and $\mathbf{b}$ directions, which corresponds to the nearest-neighbor RKKY interaction $J_H$ in Eq. \ref{H}. In order to calculate $J_H$, one can calculate the energy shift between different magnetic configurations using DFT+U simulations. For simplicity, we take CeMIn$_5$ (M=Rh, Co, Ir) as an explicit example. Experimental observations have shown that CeMIn$_5$ are AF ordered in their ground states at ambient pressure~\cite{Ida08,Sidorov02,Shang14}.
According to the lattice structure in Fig. \ref{lattice}(d), due to increasing distances between Ce atoms, the next-nearest-neighbor intra-plane exchange $J_2$ and inter-plane exchange $J_3$ are successively reduced than $J_H$ (since they decrease inversely proportional to the cube of distance), therefore, the AF phase likely takes the configuration in Fig. \ref{RKKY}(a), with an exchange energy of $E_{AF}=(-8J_H+4J_2-4J_3)S^2$ ($S=1/2$, since $f$ occupation is close to $1$). In order to directly extract $J_H$, one can flip some spins to create an intra-plane FM state (Fig. \ref{RKKY}(b)) with exchange energy of $E_{FM}=(8J_H+4J_2-4J_3)S^2$, thus the energy shift equals $\Delta E=E_{AF}-E_{FM}=-16J_HS^2$. Using DFT+U simulations for these two magnetic states (including SOC), $J_H$ can be directly evaluated. For CeM$_2$As$_2$ (M=Rh, Ir) and CeCu$_2$Si$_2$, the local spins are also staggered on $\mathbf{a}$$\mathbf{b}$ plane with a displacement of $(\mathbf{a}+\mathbf{b})/2$ between nearby Ce layers~\cite{Kibune22}, and their $J_H$ can be similarly determined. The resulting RKKY interactions of the ground states at ambient pressure are listed in Tab. \ref{tab1}, and the pressure evolutions of $J_H$ in CeRh$_2$As$_2$ and CeSb$_2$ are plotted in Fig. \ref{dlt_phase}(c) and (d), indicating an increasing of $J_H$ under pressure. Both $J_K$ and $J_H$ strengths extracted are compatible with the typical value in Ce-based materials~\cite{Fumega24}.

With the estimated Kondo and RKKY interactions, the ground-state phases of these SC materials can be qualitatively determined according to the model results in Fig. \ref{ModelCalculation}(b), and are summarized on $J_K$$J_H$ plane in Fig. \ref{dlt_phase}(d). Due to quasi-two dimensional character, the average RKKY exchange energy per Ce atom equals $2J_HS^2$, thus the energy ratio between Kondo and RKKY interactions can be measured by $J_K/2J_H$~\cite{Doniach77}, which are given in Tab. \ref{tab1}.
For materials with weak Kondo and strong RKKY interactions (CeSb$_2$ and CeCu$_2$Si$_2$), RKKY interaction dominates, leading to long-range magnetic correlation hence AF ordered phase; for strong Kondo and weak RKKY interactions (CeIr$_2$As$_2$, pressured CeRh$_2$As$_2$ and CeSb$_2$), RKKY is overwhelmed by Kondo coupling, leading to KP phase with vanished magnetic correlation.
For CeRh$_2$As$_2$ (at ambient pressure) and CeMIn$_5$ (M=Rh, Co, Ir), the magnitudes of $J_K/2J_H$ are close to $1$, indicating that the magnetic correlation and Kondo hybridization are compatible, and their interplay and possible coexistence make the ground states close to the narrow AF+SC coexisting region. Explicitly, in CeRhIn$_5$, the medium RKKY slightly overcomes the medium-low Kondo coupling, leading to weak AF order near AF+SC region; for medium RKKY and medium Kondo coupling in CeRh$_2$As$_2$, both interactions dominate, leading to AF+SC coexisting phase; while for CeCoIn$_5$ and CeIrIn$_5$ with medium RKKY and strong Kondo interactions, Kondo hybridization coexists with residual short-range magnetic correlation, resulting in paramagnetic heavy-fermion SC phase. It should be noted that in the model results in Fig. \ref{ModelCalculation}(b), $J_K$ and $J_H$ are in unit of $t$, and also due to the simplification of the effective model (neglecting of three-dimensional structure and inter-plane exchanges in these materials, etc), for CeRh$_2$As$_2$ (at ambient pressure) and CeMIn$_5$ (M=Rh, Co, Ir) which have compatible $J_K$ and $J_H$, the phases may not be precisely located. In spite, their relative positions on the evolution path (Fig. \ref{ModelDescription}(d)) can be correctly determined, see Fig. \ref{dlt_phase}(d).

As shown in Fig. \ref{dlt_phase}(c), volume compression significantly enhances both Kondo and RKKY strengths~\cite{Doniach77} in CeRh$_2$As$_2$ and CeSb$_2$, nevertheless, the growth of Kondo coupling seems to be more rapid than RKKY. During the competition, Kondo interaction gradually overwhelms RKKY interaction and dominates at high pressure, consequently the magnetic correlations between Ce atoms are suppressed. The evolutions of $J_K$ and $J_H$ under compression are indicated by green and red arrows for CeRh$_2$As$_2$ and CeSb$_2$ in Fig. \ref{dlt_phase}(d), respectively. By comparison with Fig. \ref{ModelCalculation}(b), due to much slower enhancement of $J_H$ than $J_K$ under pressure, the phase evolutions follow similar path in Fig. \ref{ModelCalculation}(a), i.e., from AF to AF+SC, then to SC phase, starting from different ground states at ambient pressure (AF and AF+SC for CeSb$_2$ and CeRh$_2$As$_2$, respectively). Moreover, further increase of pressure causes the localized-itinerant crossover at critical pressure $P_c$ at which the $f$ electrons become fully itinerant, as a consequence, the magnetic correlation $\Delta$ between $f$ electrons eventually vanishes at $P_c$ due to overwhelming $J_K$ over $J_H$, hence destructs the heavy-fermion SC state and induces a further SC-to-KP transition at $P_c$. The SC-to-KP transition can be qualitatively located near the Kondo-coherent CeIr$_2$As$_2$, 6$\%$ compressed CeRh$_2$As$_2$ and near 20$\%$ compressed CeSb$_2$, at which they undergo localized-itinerant transitions, see the purple line in Fig. \ref{dlt_phase}(d).
Since the $f$ electrons are simply described by local moments in the effective model, the localized-itinerant crossover and corresponding SC-KP transition may not be directly obtained in the model level, nevertheless, such SC transition can still be understood by the ignorable SC order $\Delta_{\mathrm{SC}}$ at large $J_K$ (see Fig. \ref{ModelCalculation}(a), $\Delta_{\mathrm{SC}}$ decreases rapidly above $J_K$=2.6t), on the right side of SC phase in Fig. \ref{ModelCalculation}(b).

Based on above analysis of the effective-model and first-principle calculations, we have depicted a qualitative description regarding the commonality in phase-evolution processes of heavy-fermion SC compounds under pressure. In general, with increasing pressure, these compounds follow similar path in Fig. \ref{ModelDescription}(d), however, they start at distinct ground-state phases at ambient pressure, depending on the competition between Kondo and RKKY interactions. With increasing pressure, two notable transitions take place successively, one is the magnetic QCP, the other is the SC-KP transition. The QCP corresponds to the AF transition which separates the SC states into AF+SC coexisting phase and paramagnetic SC phase (see the yellow line in Fig. \ref{dlt_phase}(d)), while the SC transition is induced by the localized-itinerant crossover of $f$ electrons, which eliminates the magnetic correlation hence destructs heavy-fermion SC state and produces a KP phase thereafter.

\section{conclusion and discussion}

To summarize, based on comprehensive studies of typical Ce-based superconductors via combining first-principle simulations with effective-model calculations, we have presented a proper description regarding their paths of phase evolutions under pressure. Particularly, we have demonstrated that at the presence of $c$-$f$ hybridization, the short-range singlet pairing between $f$ electrons in the context of AF long-range order can drive a notable AF+SC coexisting phase, which is separated with paramagnetic SC phase by a magnetic transition, thus gives a natural explanation to the observed magnetic QCP inside the SC phases. Furthermore, the crossover from localization to itineration for $f$ electrons under increasing pressure gives a theoretical interpretation for the SC-to-KP transition in these SC compounds at high pressure. By examining the degree of Kondo coupling and RKKY superexchange in these materials, we have eventually depicted a schematic phase diagram with regard to the pressure dependence of their ground-state phases from ambient pressure. Our description for the ground states of heavy-fermion SC systems are consistent with the phenomenological two-fluid theory~\cite{Yang15,Yang152}, and may help to achieve a microscopic explanation for the pressure-temperature phase diagram.

Although our method gives an appropriate description regarding the main feature of the phase evolutions in typical Ce-based heavy-fermions SC materials under pressure, the explicit processes may differ, e.g., other than $d$-wave symmetry~\cite{An10,Akbari12}, the SC pairing symmetries can be $s_{\pm}$-wave~\cite{Ikeda15} or nodeless mixing type $d$-wave~\cite{Pang18}.
Besides, in CeRh$_2$As$_2$ and other noncentrosymmetric SC compounds~\cite{Bauer04}, the mixing of spin-singlet and triplet pairings may induce the transition or mixing between even and odd parities under pressure~\cite{Siddiquee22}. In addition, under higher pressure above the SC transition point, a distinct SC state can arise in CeCu$_2$Si$_2$, which may be induced by valence-electron fluctuation or orbital transition~\cite{Yuan03,Pourovskii14}. Moreover, beside the SC pairing mediated by magnetic correlation, other SC pairing mechanisms can also work, such as Kondo-destruction QCP~\cite{Gegenwart08} and quadrupolar-exciton-mediated pairing in PrOs$_4$Sb$_{12}$~\cite{Aoki07}. Therefore, the explicit theoretical explanations for the pressure-dependence of these distinct heavy-fermion SC systems require extended studies beyond our minimal effective model. Our study can also help to understand the heavy-fermion SC states emerging around the ferromagnetic quantum transition point in UCoGe and UGe$_2$~\cite{Slooten09,Huxley03}.

\section{appendix I: Detail of Model calculation}

Based on the mean-field schedule described in Sec. III, and in consideration of the longitudinal polarization terms in Kondo coupling and RKKY interaction, the model Hamiltonian is written in momentum space as
\begin{align}
&\mathcal{H}=\sum_{\mathbf{k},\sigma}\epsilon_{\mathbf{k}}c_{\mathbf{k}\sigma}^{\dag}c_{\mathbf{k}\sigma}
+\sum_{\mathbf{k},\sigma}\lambda f_{\mathbf{k}\sigma}^{\dag}f_{\mathbf{k}\sigma}\nonumber\\
&+\sum_{\mathbf{k}}\Delta_{\mathbf{k}}(f_{\mathbf{k}\uparrow}^{\dag}f_{-\mathbf{k} \downarrow}^{\dag}+h.c.) \nonumber\\
&+\frac{1}{2}J_{K\parallel}m_f{\sum_{\mathbf{k} \sigma}}\sigma{c_{\mathbf{k}+\mathbf{Q}\sigma}^{\dag}}c_{\mathbf{k}\sigma}-m\sum_{\mathbf{k}\sigma}\sigma f_{\mathbf{k}+\mathbf{Q}\sigma}^{\dag}f_{\mathbf{k}\sigma} \nonumber\\
&-\frac{1}{4}J_{K\perp}{\sum_{\mathbf{k}\sigma}}(3V_sf_{\mathbf{k}\sigma}^{\dag}c_{\mathbf{k}\sigma}-V_t\sigma c_{\mathbf{k}\sigma}^{\dag}f_{\mathbf{k}+\mathbf{Q}\sigma}+h.c.) \nonumber\\
&+N[\frac{1}{2}J_{K\perp}(3V_{s}^2-V_{t}^2)+J_{H\perp}{\Delta}^2+J_{K\parallel}m_c{m_f}\nonumber\\
&+2J_{H\parallel}m_{f}^2-\lambda],
\label{Hmean}\end{align}
where $\Delta_{\mathbf{k}}=-J_{H\perp}\Delta(\cos{\mathbf{k}_x}-\cos{\mathbf{k}_y})$, $m=2J_{H\parallel}m_f+\frac{1}{2}J_{K\parallel}m_c$, and $N$ is the total number of lattice sites. In above equation, the strengths of transverse channels ($J_{K\perp}$, $J_{H\perp}$) and longitudinal channels ($J_{K\parallel}$, $J_{H\parallel}$) in both Kondo coupling and RKKY interaction are treated independently for better performance. By defining a Nambu operator
$
\mathbf{\Phi}_\mathbf{k}=(
c_{\mathbf{k}\uparrow}c_{\mathbf{k}+\mathbf{Q}\uparrow}c_{-\mathbf{k}\downarrow}^{\dag}c_{-\mathbf{k}+\mathbf{Q}\downarrow}^{\dag} f_{\mathbf{k}\uparrow}
f_{\mathbf{k}+\mathbf{Q}\uparrow}f_{-\mathbf{k}\downarrow}^{\dag}f_{-\mathbf{k}+\mathbf{Q}\downarrow}^{\dag}
)^T
$, the mean-field Hamiltonian Eq. \ref{Hmean} is rewritten in a compact form as
$
\mathcal{H}=N\eta+\sum_{\mathbf{k}}{\bf\Phi}^{\dag}_{\mathbf{k}}\mathbf H_{\mathbf k}\bf\Phi_{\mathbf{k}}
$, in which the summation of $\mathbf{k}$ is now restricted in the AF magnetic Brillouin zone, with
$
\eta=\frac{1}{2}J_{K\perp}(3V_{s}^2-V_{t}^2)+J_{H\perp}{\Delta}^2
+J_{K\parallel}m_c{m_f}+2J_{H\parallel}m_{f}^2-\mu+\mu n_c
$, and the Hamiltonian matrix
\begin{align}
\mathbf{H}_{\mathbf{k}}=
\left(
\begin{array}{cc}
\mathbf{A}_\mathbf{k} & \mathbf{V}\\
\mathbf{V} & \mathbf{B}_\mathbf{k}
\end{array}
\right)
\label{Hk},\end{align}
with
\begin{align}
\mathbf{A}_{\mathbf{k}}=
\left(
\begin{array}{cccc}
\epsilon_\mathbf{k} & \frac{1}{2}J_{K\parallel}m_f & 0 & 0\\
\frac{1}{2}J_{K\parallel}m_f & \epsilon_{\mathbf{k}+\mathbf{Q}} & 0 & 0\\
0 & 0 & -\epsilon_\mathbf{k} & \frac{1}{2}J_{K\parallel}m_f\\
0 & 0 & \frac{1}{2}J_{K\parallel}m_f & -\epsilon_{\mathbf{k}+\mathbf{Q}}
\end{array}
\right)
,\end{align}
\begin{align}
\mathbf{V}=
\left(
\begin{array}{cccc}
 -\frac{3}{4}J_{K\perp}V_s & \frac{1}{4}J_{K\perp}V_t & 0 & 0\\
 \frac{1}{4}J_{K\perp}V_t & -\frac{3}{4}J_{K\perp}V_s & 0 & 0\\
 0 & 0 & \frac{3}{4}J_{K\perp}V_s & \frac{1}{4}J_{K\perp}V_t \\
 0 & 0 & \frac{1}{4}J_{K\perp}V_t & \frac{3}{4}J_{K\perp}V_s
\end{array}
\right)
,\end{align}
\begin{align}
\mathbf{B}_{\mathbf{k}}=
\left(
\begin{array}{cccc}
 \lambda & -m & \Delta_\mathbf{k} & 0 \\
 -m & \lambda & 0 & -\Delta_\mathbf{k} \\
 \Delta_\mathbf{k} & 0 & -\lambda & -m \\
 0 & -\Delta_\mathbf{k} & -m & -\lambda
\end{array}
\right)
.\end{align}
\begin{figure*}[tbp]
\hspace{-0cm} \includegraphics[totalheight=2.7in]{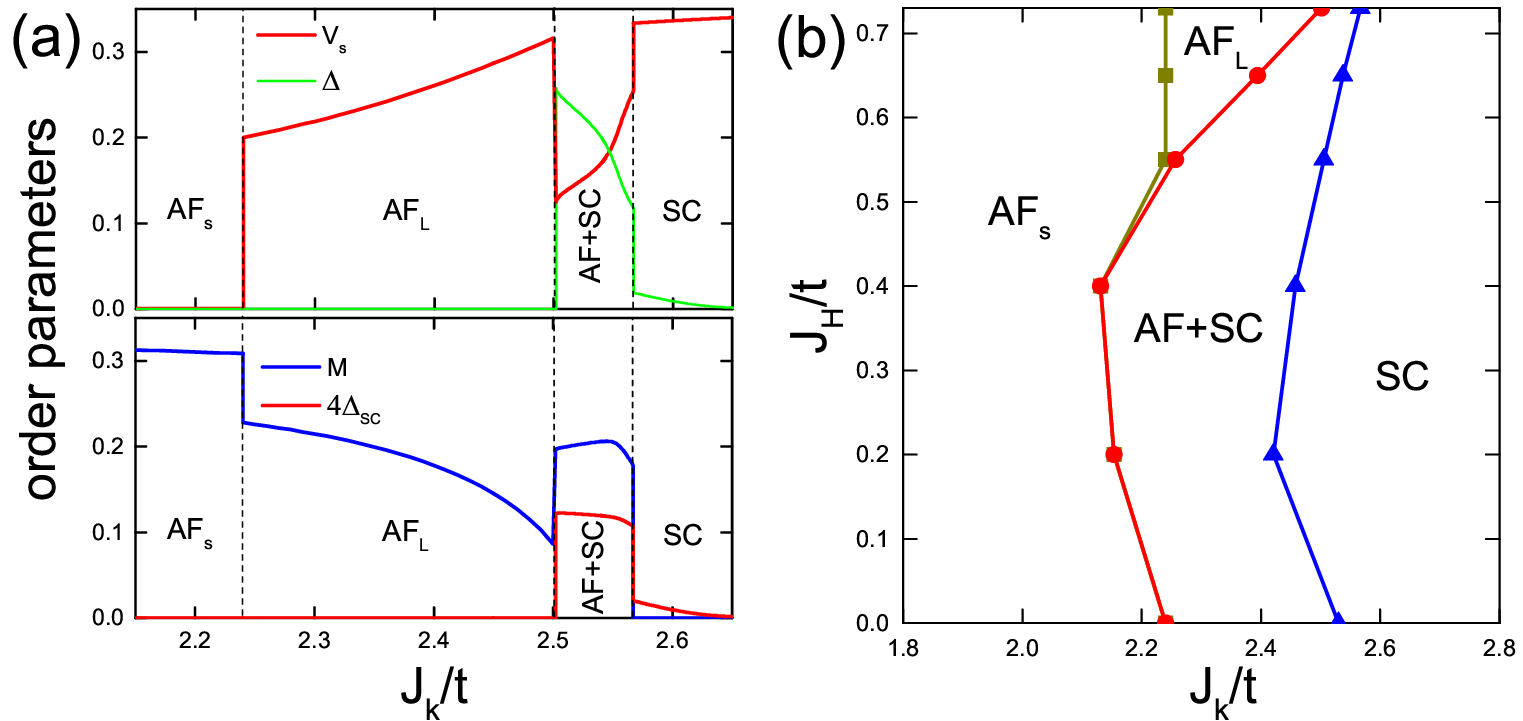}
\caption{
(a) Variation of $c$-$f$ hybridization $V_s$, $f$-$f$ singlet pairing $\Delta$, staggered magnetization $M$, and SC order $\Delta_{sc}$ as functions of Kondo coupling $J_K$. (b) Phase diagram of the effective model for heavy-fermion SC system. Parameters are set as $J_{K\parallel}=0.86 J_{K\perp}$, $J_{H\parallel}=0$, $t^\prime=0.1$, and $n_c=0.95$. In (a), $J_{H\perp}=0.73$. On the horizontal and vertical axes, the magnitudes of $J_K$ and $J_H$ are defined according to the transverse parts $J_{K\perp}$ and $J_{H\perp}$, respectively. All energies are in units of nearest-neighbor hopping strength $t$ between conduction electrons.
}
\label{ModelCalculation}
\end{figure*}
In general, the quasi-particle spectrums have to be obtained by numerical diagonalization of $\mathbf{H}_{\mathbf{k}}$ via $\mathbf{U}_{\mathbf k}^{\dag}\mathbf{H}_{\mathbf k}\mathbf{U}_{\mathbf k}={\Lambda}_\mathbf{k}=\mathrm{diag}(E^{(1)}_{\mathbf{k}},E^{(2)}_{\mathbf{k}},...,E^{(8)}_{\mathbf{k}})
$, in which $\mathbf{U}_{\mathbf{k}}$ is a unitary matrix, and can be extracted through numerical diagonalization. Therefore, by transforming Nambu operator $
\mathbf\Phi_\mathbf{k}$ to the quasi-particle operator $\mathbf\Psi_ \mathbf{k}$ by $\mathbf\Phi_\mathbf{k}=\mathbf U_\mathbf{k}\mathbf\Psi_ \mathbf{k}$, with $\mathbf\Psi_{\mathbf{k}}^{\dag}=(\mathbf\Psi_{\mathbf{k}}^{(1)\dag},...,\mathbf\Psi_{\mathbf{k}}^{(8)\dag})$,
the matrix expression of Hamiltonian is written as $
\mathbf\Phi_\mathbf{k}^{\dag}\mathbf H_{ \mathbf{k}}\mathbf\Phi_\mathbf{k}=\mathbf\Psi_{\mathbf{k}}^{\dag}\mathbf\Lambda_{\mathbf{k}}\mathbf\Psi_{\mathbf{k}}
$, in a diagonalized form. Now one can evaluate the ground-state expectation values for quadratic products of $d$ and $f$ operators through ($n,m=1,...,8$)
\begin{align}\langle nm\rangle_{\mathbf{k}}\equiv&\langle (\Phi^\dag_{\mathbf{k}})_n(\Phi_{\mathbf{k}})_m\rangle\nonumber\\
=&\sum_{i,j}(\mathrm{U}_{\mathbf{k}})^*_{ni}(\mathrm{U}_{\mathbf{k}})_{mj}\langle\Psi^{(i)\dag}_{\mathbf{k}}\Psi^{(j)}_{\mathbf{k}}\rangle\nonumber\\
=&\sum_{i}(\mathrm{U}_{\mathbf{k}})^*_{ni}(\mathrm{U}_{\mathbf{k}})_{mi}\Theta(-E^{(i)}_{\mathbf{k}}).
\label{nm}\end{align}
Consequently, the self-consistent equations determining the chemical potential $\mu$, Lagrangian multiplier $\lambda$, and mean-field parameters $\Delta$, $V_s$, $V_t$, $m_c$, $m_f$ can be derived via
\begin{align}
n_c&=\frac{1}{N}\sum_{\mathbf{k} \sigma}\langle c_{\mathbf{k} \sigma}^{\dag}c_{\mathbf{k} \sigma}+c_{\mathbf{k}+\mathbf{Q} \sigma}^{\dag}c_{\mathbf{k}+\mathbf{Q} \sigma}\rangle,\label{eq1}\end{align}
\begin{align}
1&=\frac{1}{N}\sum_{\mathbf{k} \sigma}\langle f_{\mathbf{k} \sigma}^{\dag}f_{\mathbf{k} \sigma}+f_{\mathbf{k}+\mathbf{Q} \sigma}^{\dag}f_{\mathbf{k}+\mathbf{Q} \sigma}\rangle,\end{align}
\begin{align}
m_c&=-\frac{1}{2N}\sum_{\mathbf{k} \sigma}\sigma\langle c_{\mathbf{k} \sigma}^{\dag}c_{\mathbf{k}+\mathbf{Q}\sigma}+c_{\mathbf{k}+\mathbf{Q} \sigma}^{\dag}c_{\mathbf{k}\sigma}\rangle,\end{align}
\begin{align}
m_f&=\frac{1}{2N}\sum_{\mathbf{k} \sigma}\sigma\langle f_{\mathbf{k} \sigma}^{\dag}f_{\mathbf{k}+\mathbf{Q}\sigma}+f_{\mathbf{k}+\mathbf{Q} \sigma}^{\dag}f_{\mathbf{k}\sigma}\rangle,\end{align}
\begin{align}
\Delta&=\frac{1}{2N}\sum_{\mathbf{k}}(\cos \mathbf{k}_x-\cos \mathbf{k}_y)
\langle
f^{\dag}_{\mathbf{k}\uparrow}f^{\dag}_{-\mathbf{k}\downarrow}
-f^{\dag}_{-\mathbf{k}\downarrow}f^{\dag}_{\mathbf{k}\uparrow}\nonumber\\
&-f^{\dag}_{\mathbf{k}+\mathbf{Q}\uparrow}f^{\dag}_{-\mathbf{k}+\mathbf{Q}\downarrow}
+f^{\dag}_{-\mathbf{k}+\mathbf{Q}\downarrow}f^{\dag}_{\mathbf{k}+\mathbf{Q}\uparrow}
\rangle,\end{align}
\begin{align}
\mathbf {V}_s&=\frac{1}{2N}\sum_{\mathbf{k}}
\langle
c^{\dag}_{\mathbf{k}\uparrow}f_{-\mathbf{k}\uparrow}
+c^{\dag}_{\mathbf{k}+\mathbf{Q}\uparrow}f_{\mathbf{k}+\mathbf{Q}\uparrow}\nonumber\\
&+c^{\dag}_{-\mathbf{k}\downarrow}f_{-\mathbf{k}\downarrow}
+c^{\dag}_{-\mathbf{k}+\mathbf{Q}\downarrow}f_{-\mathbf{k}+\mathbf{Q}\downarrow}
\rangle,\end{align}
\begin{align}
\mathbf {V}_t&=\frac{1}{2N}\sum_{\mathbf{k}}
\langle
c^{\dag}_{\mathbf{k}\uparrow}f_{\mathbf{k}+\mathbf{Q}\uparrow}
+c^{\dag}_{\mathbf{k}+\mathbf{Q}\uparrow}f_{\mathbf{k}\uparrow}\nonumber\\
&-c^{\dag}_{-\mathbf{k}\downarrow}f_{-\mathbf{k}+\mathbf{Q}\downarrow}
-c^{\dag}_{-\mathbf{k}+\mathbf{Q}\downarrow}f_{-\mathbf{k}\downarrow}
\rangle,\label{eq7}\end{align}
by substituting Eq. \ref{nm}. Note that $\mathbf{k}$-summations in above equations are confined in the magnetic Brillouin zone. Eqs. \ref{eq1}-\ref{eq7} are then solved self-consistently through numerical iterations.

In Sec. II, we have demonstrated that pressure can remarkably enhance the Kondo coupling strength in heavy-fermion SC compounds, thus based on the effective model, we first calculate the variation of order parameters as functions of Kondo coupling strength $J_K$, and the results are illustrated in Fig. \ref{ModelCalculation}(a). Under proper magnitude of RKKY interaction $J_H$, the Kondo hybridization $V_s$, staggered magnetization $M=m_f-m_c$, and SC order parameter $\Delta_{sc}$ evolute along a representative path. At weak Kondo coupling $J_K$, long-range magnetic correlation dominates meanwhile the Kondo hybridizations $V_s$ and $V_t$ vanish, leading to an AF phase ($M\neq0$) with small Fermi surface occupying only by conduction electrons (denoted by AF$_s$). While $2.24t<J_K<2.502t$, $c$-$f$ hybridizations set in ($V_s,V_t\neq0$) and coexist with AF long-range order, leading to another AF phase with large Fermi surface constructed by both $c$ and $f$ electrons (denoted by AF$_L$). With strong Kondo coupling ($J_K>2.567t$), the AF long-range order is fully suppressed, however, the short-range magnetic correlation $\Delta$ survives and coexists with Kondo hybridization $V_s$, driving a paramagnetic $d$-wave SC state with non-vanishing SC order $\Delta_{sc}\neq0$, in which the SC pairing is caused by heavy quasiparticles combining $c$ with $f$ electrons~\cite{Liu12}. Notably, in intermediate region of Kondo coupling ($2.502t<J_K<2.567t$), a novel phase emerges, which coexists AF order ($M\neq0$, meanwhile the short-range magnetic correlation persists) with heavy-fermion SC pairing ($V_s,\Delta_{sc}\neq0$), and can be ascribed to the AF+SC phases observed in some Ce-based heavy-fermion materials such as CeCu$_2$Si$_2$, CeRhIn$_5$, CeSb$_2$ and CeRh$_2$As$_2$, under ambient or high pressures~\cite{Yuan03,Ida08,Squire22,Siddiquee22}.

In summation of the evolution process with Kondo coupling strength, three notable transitions arise as a consequence of competition between Kondo hybridization and magnetic correlation, and the resulting phase diagram is plotted on $J_K$-$J_H$ plane in Fig. \ref{ModelCalculation}(b). Firstly, the enhancement of Kondo hybridization $V_s$ (under increasing $J_K$) reduces the long-rang AF order and drives a magnetic transition denoted by blue solid line in Fig. \ref{ModelCalculation}(b), which divides the SC states into AF+SC phase and paramagnetic SC phase, respectively. Although long-range AF order is already diminished in the paramagnetic SC state, short-range magnetic correlation $\Delta$ survives and drives the paramagnetic heavy-fermion SC state in company with Kondo hybridization. Secondly, while Kondo coupling $J_K$ is further reduced in AF+SC phase, the long-range AF order gains lower energy than the short-range singlet correlation, hence a SC transition occurs at which $\Delta$ vanishes, denoted by red line in Fig. \ref{ModelCalculation}(b). Continual decreasing of Kondo coupling reduces the Kondo hybridization $V_s$, and eventually produces a Fermi-surface reconstruction in the AF states, from large to small Fermi surfaces at which $V_s$ vanishes, indicated by the dark yellow line in Fig. \ref{ModelCalculation}(b). It should be noted that the present calculation gives a first-order AF transition between AF+SC phase and paramagnetic SC phase in Fig. \ref{ModelCalculation}(b), owing to our setting of model parameters which deviates from particle-hole symmetry in order to make the transition more noticeable. Such first-order magnetic transition is similar to that observed in CeIn$_3$ under pressure~\cite{Kawasaki08}, besides, the magnetic transition can be continuous one under particle-hole symmetric setting of model parameters~\cite{Li15}, which can give a explanation to the magnetic QCP of CePd$_2$Si$_2$ and CeCu$_2$Si$_2$ under pressure~\cite{Arndt11}.

\acknowledgments
H. Li and Y. Liu are grateful to earlier guidance and collaboration with G. M. Zhang. This work is supported by National Natural Science Foundation of China (No. 12364023 and 12264011), GuikeAD20159009, the National Key Research and Development Program of China (No. 2021YFB3501503), and the Foundation of LCP.

\end{document}